\documentclass[twocolumn, twocolappendix]{aastex631}

\usepackage{graphicx}	
\usepackage{amsmath}	
\usepackage{bm}
\usepackage{newtxtext,newtxmath}
\usepackage[T1]{fontenc}
\usepackage{comment}
\usepackage[normalem]{ulem}
\usepackage{CJK}

\DeclareRobustCommand{\VAN}[3]{#2}
\let\VANthebibliography\thebibliography
\def\thebibliography{\DeclareRobustCommand{\VAN}[3]{##3}\VANthebibliography}

\def\s{\mathrm{s}} 
\def\millis{\mathrm{m}\s} 

\def\days{\mathrm{d}} 

\def\GHz{\mathrm{GHz}} 

\def\m{\mathrm{m}} 
\def\cm{\mathrm{cm}} 
\def\km{\mathrm{km}} 

\def\g{\mathrm{g}} 

\def\erg{\mathrm{erg}} 
\def\J{\mathrm{J}} 
\def\K{\mathrm{K}} 

\def\ergsec{\erg/\s} 

\def\rad{\mathrm{rad}} 

\def\G{{\rm G}} 
\def\T{{\rm T}} 

\def\ej{\mathrm{ej}}

\def\obs{\mathrm{obs}} 
\def\max{\mathrm{max}} 
\def\sh{\mathrm{sh}} 
\def\stop{\mathrm{stop}} 
\def\BM{\mathrm{BM}} 
\def\ST{\mathrm{ST}} 
\def\R{\mathrm{R}} 
\def\TR{\mathrm{TR}} 
\def\NR{\mathrm{NR}} 
\def\qelectron{q_\mathrm{e}} 
\def\me{m_\mathrm{e}} 
\def\mprot{m_\mathrm{p}} 
\def\epsilone{\epsilon_\mathrm{e}} 
\def\epsilonB{\epsilon_\mathrm{B}} 
\def\peak{\mathrm{peak}} 

\newcommand{\di}{\partial}

\renewcommand{\vec}{\mathbf}
\renewcommand{\epsilon}{\varepsilon}

\newcommand{\const}{\mathrm{const.}}

\newcommand{\engine}{\mathrm{e}}
\newcommand{\te}{t_\mathrm{e}}
\newcommand{\td}{t_\mathrm{d}}

\newcommand{\msun}{\mathrm{M}_\odot}

\newcommand{\jet}{\mathrm{j}}
\renewcommand{\vec}{\mathbf}
\renewcommand{\epsilon}{\varepsilon}

\newcommand{\de}{\mathrm{d}}

\renewcommand{\vec}{\mathbf}
\renewcommand{\epsilon}{\varepsilon}
\renewcommand{\vec}[1]{\bm{#1}}

\newcommand{\tp}[1]{\textcolor{blue}{#1}}

\begin{document}
\begin{CJK}{UTF8}{gbsn}

\title{Simulating short GRB jets in realistic late binary neutron star merger environments}

\author[0000-0002-4243-3889]{Matteo Pais}
\affiliation{Racah Institute for Physics, The Hebrew University, Jerusalem 91904, Israel}
\affiliation{INAF, Osservatorio Astronomico di Padova, Vicolo dell'Osservatorio 5, I-35122, Padova, Italy}

\author[0000-0002-7964-5420]{Tsvi Piran}
\affiliation{Racah Institute for Physics, 
The Hebrew University, Jerusalem 91904, Israel}

\author[0000-0003-4988-1438]{Kenta Kiuchi (木内 建太)}
\affiliation{Max Planck Institute for Gravitational Physics (Albert Einstein Institute) \\ 
Am M\"{u}hlenberg 1, 14476, Potsdam-Golm, Germany}
\affiliation{Center for Gravitational Physics and Quantum Information, \\
Yukawa Institute for Theoretical Physics, Kyoto University,
606-8502, Kyoto, Japan}

\author[0000-0002-4979-5671]{Masaru Shibata (柴田 大)}
\affiliation{Max Planck Institute for Gravitational Physics (Albert Einstein Institute) \\ 
Am M\"{u}hlenberg 1, 14476, Potsdam-Golm, Germany}

\affiliation{Center for Gravitational Physics and Quantum Information, \\
Yukawa Institute for Theoretical Physics, Kyoto University,
606-8502, Kyoto, Japan}

\begin{abstract}
The electromagnetic emission and the afterglow observations of the binary neutron star merger event GW 170817A confirmed the association of the merger with a short gamma-ray burst (sGRB) harboring a narrow ($5$--$10^\circ$) and powerful ($10^{49}$--$10^{50}~\erg$) jet. 
Using the 1~second-long neutrino-radiation-GR-MHD simulation of coalescing neutron stars of \citet{kiuchi_self-consistent_2023}, and following the semi-analytical estimates of \citet{pais_collimation_2023} we inject a narrow, powerful, unmagnetized jet into the post-merger phase. We explore different opening angles, luminosities, central engine durations, and times after the merger. 
We explore early ($0.1~\s$ following the merger) and late ($1~\s$) jet launches; the latter is consistent with the time delay of $\approx 1.74~\s$ observed between GW 170817 and GRB 170817A.
We demonstrate that the semi-analytical estimates correctly predict the jets' breakout and collimation conditions.
When comparing our synthetic afterglow light curves to the observed radio data of GW170807, we find a good agreement for a $3 \times 10^{49}$ ergs jet launched late with an opening angle in the range $\simeq 5^\circ$--$7^\circ$. 

\end{abstract}

\keywords{stars: jets -- gamma-ray burst: general --  stars: neutron -- hydrodynamics}



\section{Introduction}

The merger of a neutron star binary (BNS) or a neutron star and a black hole (BH) is the most promising candidate for the detection of gravitational waves and their electromagnetic counterparts:  a short GRB \citep{eichler_nucleosynthesis_1989}, a kilonova \citep{li_transient_1998, kulkarni_modeling_2005, metzger_electromagnetic_2010} and their afterglows. 
The tentative observations of kilonovae following sGRB 130603B \citep{tanvir_kilonova_2013,berger_r-process_2013} and other sGRB \citep[e.g.,][]{yang_possible_2015, jin_macronova_2016, jin_light_2015, gompertz_diversity_2018, ascenzi_luminosity_2019,  jin_kilonova_2020, lamb_short_2019, rossi_comparison_2020, fong_broadband_2021} suggested such association but these were not a sufficient affirmation. 
The BNS merger GW 170817 provided the first evidence of the detection of a gravitational wave from a BNS merger \citep{abbott_GW170817_2017}. It was followed by a multi-wavelength electromagnetic signal \citep{abbott_multi-messenger_2017}. 

Numerous numerical simulations of mergers have been carried out over the years \citep[e.g.,][]{davies_merging_1994,ruffert_coalescing_1995, rosswog_mass_1998, shibata_fully_1999, shibata_simulation_2000, rosswog_merging_2000, oechslin_conformally_2002, Shibata:2005ss, Shibata:2006nm,  anderson_simulating_2008, liu_general_2008, giacomazzo_can_2009, giacomazzo_accurate_2011, giacomazzo_formation_2013, piran_electromagnetic_2013, bauswein_systematics_2013, rosswog_multimessenger_2013, hotokezaka_mass_2013, kiuchi_high_2014,  dionysopoulou_general-relativistic_2015, giacomazzo_producing_2015, Sekiguchi:2015dma, sekiguchi_dynamical_2016, foucart_impact_2016, kawamura_binary_2016, radice_dynamical_2016, ruiz_binary_2016, ciolfi_general_2017, dietrich_gravitational_2017-1, ruiz_general_2017, kiuchi_global_2018, ruiz_jet_2018, ciolfi_first_2019, ruiz_effects_2019, aguilera-miret_turbulent_2020, ciolfi_collimated_2020, mosta_magnetar_2020, ruiz_magnetohydrodynamic_2020, Hayashi:2021oxy, ruiz_jet_2021, aguilera-miret_universality_2022, palenzuela_turbulent_2022, sun_jet_2022, aguilera-miret_role_2023, combi_grmhd_2023, combi_jets_2023, most_flares_2023, kiuchi_large-scale_2024}. 
State-of-the-art simulations rely on fully general relativistic numerical schemes, approximated neutrino transport, and a realistic equation of state for the neutron stars. 
Before coalescence, the system starts to eject mass because of the tidal forces. Typically, the resulting ejecta expands at $0.1~$c \citep{hotokezaka_mass_2013}, with a small fraction that moves faster, around $0.6~$c or even faster \citep{kiuchi_sub-radian-accuracy_2017, fujibayashi_comprehensive_2023}.  
A second source of ejecta is the bound material torn from the NS after the merger with enough angular momentum to form a thick torus around the newly created compact object from the fusion. 
Magneto-rotational instability (MRI) generates viscosity, leading to accretion \citep{zenati_bound_2023, zenati_dynamics_2024}.
A viscous-driven wind emerges from the disk and carries a fraction of the disk material at low ($<0.1~$c) velocities \citep[e.g.,][]{fernandez_delayed_2013, Just:2014fka, Fujibayashi:2020qda}.

The afterglow observations of GW 170807 revealed that it involved a relativistic jet launched at an angle of about $18^\circ$ away from us, which means that while we did not see the direct emission from this jet, as observed in a typical GRB, a powerful relativistic jet was present in this system \citep[e.g.,][]{troja_x-ray_2017, hallinan_radio_2017, mooley_mildly_2018, mooley_superluminal_2018, ghirlanda_compact_2019}. 
Close examination of the $\gamma-$ray observations and the afterglow shows that GRB 170817A was not a regular sGRB. 
Its total isotropic equivalent energy ($\sim 10^{46}~\erg$) is smaller by three orders of magnitude than the weakest sGRB measured so far \citep{gottlieb_cocoon_2018} and by four orders of magnitude than typical sGRBs \citep{nakar_short-hard_2007}.
Both the radio \citep{hallinan_radio_2017} and X-ray \citep{troja_x-ray_2017} afterglows were delayed by several days compared to a regular sGRB. 
The compactness argument \citep{kasliwal_illuminating_2017, matsumoto_constraints_2019} reveals that the observed $\gamma$-rays must have been produced in a mildly or fully relativistic outflow with a Lorenz factor of $\Gamma\simeq 2-3$. 

\begin{figure*}
    \centering
    \includegraphics[width=1.\linewidth]{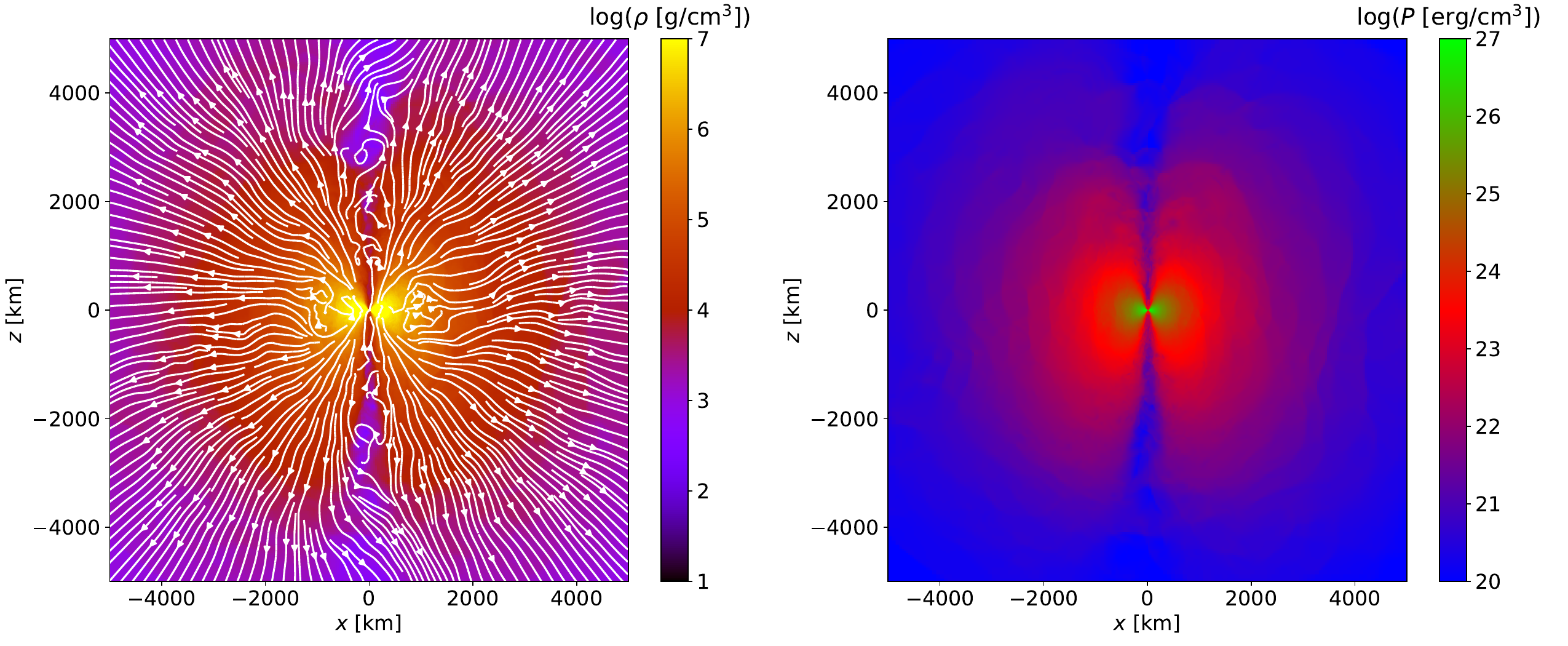}
    \vskip -0.3cm
    \caption{2D slice across the $x$-$z$ plane passing through the center of the simulation box of the initial conditions taken at $t=1.06~\s$ after the merger. 
    \textit{Left}: density slice in logarithmic units superposed to the streamlines of the velocity field. 
    \textit{Right}: pressure slice in logarithmic units. In both plots, we can notice the presence of the low-pressure/low-density polar funnel.}
    \label{fig: IC_slice}
\end{figure*}

As the jet propagates through the expanding ejecta, it forms a high-pressure bubble (the cocoon) that engulfs the jet \citep[e.g.,][]{bromberg_are_2011}. 
A natural mechanism that could have produced the observed $\gamma$-rays in GRB 170817A is a shock breakout of the cocoon from the engulfing ejecta \citep{nakar_observable_2017, kasliwal_illuminating_2017, gottlieb_cocoon_2018, beloborodov_relativistic_2020}, while the interaction of the jet and cocoon with the circumburst medium produced the afterglow.

The propagation of the jet in a surrounding medium has been studied by numerous authors both analytically \citep[e.g.,][]{blandford_twin-exhaust_1974, begelman_overpressured_1989, meszaros_tev_2001, matzner_supernova_2003, Lazzati_universal_2005, bromberg_are_2011, hamidani_jet_2019, hamidani_jet_2021, hamidani_cocoon_2022, hamidani_cocoon_2023, garcia-garcia_semi-analytical_2024} and numerically \citep[e.g.,][]{marti_morphology_1995, marti_morphology_1997, aloy_relativistic_2000, macfadyen_supernovae_2001, zhang_propagation_2004, mizuta_collimated_2006, morsony_temporal_2007, wang_relativistic_2008, lazzati_very_2009, mizuta_angular_2009, morsony_origin_2010, nagakura_jet_2011, lopez-camara_three-dimensional_2013, ito_photospheric_2015, lopez-camara_three-dimensional_2016, harrison_numerically_2018, gottlieb_cocoon_2018, pavan_short_2021, gottlieb_propagation_2022, garcia-garcia_dynamics_2023, pavan_jet-environment_2023}.
Most of those studies were conducted for jets expanding in a uniform or spherically symmetric surrounding medium (but see \cite{Nagakura:2014hza}). Only recently, inhomogeneous post-merger environments directly imported from the outcome of a GR-MHD BNS merger simulations were considered  \citep{nativi_consequences_2020,pavan_short_2021, nativi_are_2022, pavan_jet-environment_2023}.
These studies describe how a jet dissipates energy as it propagates, enabling us to determine whether it breaks out of the surrounding matter or is choked inside.
They also show whether the surrounding pressure is sufficient to collimate the jet \citep{komissarov_magnetic_2009,  lyubarsky_asymptotic_2009}.

While uniform or spherically symmetric external matter is a good approximation for stellar envelopes (relevant for long GRBs), the ejecta of the BNS merger are far from that. 
\citet{pais_collimation_2023} explored semi-analytically the conditions for a jet to break out from the ejecta structure that arose in a BNS-post merger environment based on the latest simulation described in \citet{kiuchi_self-consistent_2023}. 
A key ingredient here is the interplay between the external pressure that acts to collimate the jet and the existence of an empty funnel along the system's rotation axis, along which the jet propagates almost freely. 
\citet{pais_collimation_2023} found that a jet with energy between $10^{48}$--$10^{50}~\erg$ and an opening angle of $3^\circ$--$7^\circ$ can break out from this ejecta. 
At the same time, they found that the breakout conditions are almost independent of the time at which they are calculated. 
In this work, we numerically test the validity of these results by injecting an unmagnetized relativistic jet in the same BNS merger environment, following its evolution numerically. 

The paper is structured as follows. 
We begin, in Section~\ref{sec: ICs}, with a brief description of the environment setup resulting from the neutrino-radiation-GRMHD simulation of a BNS merger, referring the reader to \cite{kiuchi_self-consistent_2023} for details. 
The following paragraphs briefly describe the data import procedure and the jet injection into the post-BNS environment. More details on the data structures are reported in Appendix~\ref{appendix}.
In Section~\ref{sec: results}, we describe the results of several realizations of the jet with different total energy, exploring cases of jet breakout and jet choking. 
We calculate the afterglow light curves using the energy distribution from our simulations. 
We discuss the implications of our findings for GW 170817 and other sGRBs and summarise our results in Section~\ref{sec: conclusions}.
Throughout this paper, $c$ denotes the speed of light.


\section{Setup} 
\label{sec: ICs}

\subsection{The BNS environment data}

We import our initial conditions describing the ejecta surrounding the remnant BH into~\textsc{pluto} from the fully 3D general relativistic neutrino-radiation MHD simulation of a BNS merger described in greater detail in \citet{kiuchi_implementation_2022} and \citet{kiuchi_self-consistent_2023}. 
In these simulations, the neutron star is modeled with the SFHo equation of state \citep{steiner_core-collapse_2013}, and the binary is composed of a $1.2~\mathrm{M}_\odot$ and a $1.5~\mathrm{M}_\odot$ neutron star, with a chirp mass consistent with the one observed in GW 170817. 
We employ the \textsc{lorene} library~\citep{gourgoulhon_lorene_2016} to construct a quasi-equilibrium configuration of the irrotational BNS with the initial orbital frequency of $G M_0 \Omega_0/c^3 = 0.025$, where $M_0 = 2.7~\mathrm{M}_\odot$ is the total mass, and with an orbital eccentricity of $\mathcal{O}(10^{-3})$.
The entire run lasts one full second and is the longest simulation of a BNS merger in the literature. 
The merger and the subsequent BH formation occur at $\approx 0.015~\s$ and $\approx 0.032~\s$, respectively, where $t=0$ is the start time of the simulation. 

A fraction of neutron star matter is dynamically ejected by the tidal interaction and shock heating in the merge phase and shows a fast tail with a terminal velocity up to $\simeq 0.96$~c~\citep{fujibayashi_comprehensive_2023}. A high-mass neutron star (HMNS) is formed and survives for $\sim 20~$ms after the merger before collapsing to a BH. It has a non-axisymmetric density structure and exerts a gravitational torque on the plasma, transporting angular momentum outwards. 
The torque leads to the formation of a torus of mass $\simeq 0.05~\msun$, which surrounds a newly formed spinning BH of mass $\simeq 2.55~\msun$ and a dimensionless spin of $\simeq 0.65$. 

The poloidal magnetic field is initialized with a magnetar-like strength of $10^{15}~\mathrm{G}$, and the Kelvin-Helmholtz instability at the onset of the merger amplifies it further \citep{price_producing_2006, kiuchi_high_2014, kiuchi_efficient_2015, kiuchi_global_2018, aguilera-miret_turbulent_2020, aguilera-miret_universality_2022, palenzuela_turbulent_2022, aguilera-miret_role_2023, aguilera-miret_delayed_2024, kiuchi_large-scale_2024}.
Non-axis-symmetric dynamo-driven MRI contributes to the enhancement of the magnetic field.  Still, even at this level, the magnetic field is irrelevant to the overall dynamical evolution.
The magnetic winding and radial motion of the fluid enhance the magnetic field in the torus until the electromagnetic energy saturates at $\sim 1\%$ of the internal energy of the torus.  At $t\approx 0.04~\s$ after the merger, the magnetic field energy is maximal and then decreases. At all stages, the magnetic field energy is subdominant to the total energy budget ($\sim 10^{49}~\erg$ vs $\sim 10^{51}~\erg$, see \citet{kiuchi_self-consistent_2023}). 
The subdominant magnetic field energy allows us to neglect the magnetic fields in our work, which focuses on jet propagation.

At $t=0.1~\s$, a thick torus of roughly $300~\km$ in radius surrounds the compact object, and a funnel structure has already formed with its distinctive features of low density and low pressure.
The torus expands outward due to MRI-driven turbulent viscosity transporting angular momentum, and as it expands, it cools adiabatically. 
A part of the torus matter is also ejected due to this MRI-driven turbulence as the post-merger ejecta after the neutrino cooling in the torus becomes inefficient. This occurs at $t\approx 0.3~\s$ after the merger. 
At around $t\approx 1~\s$, the post-merger ejecta has a typical velocity of around 0.1\tp{5}~c,  slower than the average value of 0.2~c reported in literature \citep[see][]{sekiguchi_dynamical_2016}. This is largely due to the post-merger ejecta that is launched at the outskirt of the massive torus formed around the black hole. This turbulent viscosity-driven post-merger ejection still lasts $\simeq 1~\s$. 

At $t=1~\s$ the thick torus surrounding the BH has a size of $\simeq 750~\km$ of radius and a comparable thickness, with an average density of $\rho_\mathrm{torus} > 10^6~\g/\cm^3$. 
A funnel with a half-opening angle of roughly $10^\circ$ and an average density below $3\times 10^3~\g/\cm^3$ exists along the polar axis. 
As the simulation progresses in time, the entire system expands in size. The total mass of the ejecta at $t=1~\s$ is approximately $1.5\times 10^{-2}\msun$.  
Fig.~\ref{fig: IC_slice} shows the initial conditions of the simulation at $t=1.06~\s$ after the merger, in which the presence of two polar funnels with low density and low pressure is noticeable. 

The post-merger secular ejecta expand in a ``numerical vacuum" of circumbinary medium with a uniform pressure of $1.35 \times 10^{12}~\erg/\cm^3$ (in order to avoid spurious random motion) and a density of $\rho \simeq 0.17~\g/\cm^3$. 
This results in a mass surrounding the merger of approximately $1.2 \times 10^{-3}~\msun$ enclosed in a radius of $R_{\mathrm{max}} = 1.5 \times 10^5~\km$, which is an order of magnitude less than the mass of the ejecta. 
The fast tail of the dynamical ejecta stretching to 0.96\,c gradually slows down due to interaction with the external density that occupies the "numerical vacuum'' (see \citet{kiuchi_self-consistent_2023} and \citet{fujibayashi_comprehensive_2023} for more details) leaving behind only slower, mildly-relativistic material.
Apart from influencing the low-mass, the very high-velocity tail of the unbound material, this artificial vacuum doesn't significantly impact the ejecta's expansion within this simulation box. Hence,  it does not influence our results.

\subsection{Computational setup}

We construct our 2.5D initial conditions from the BNS simulation by cylindrically averaging the original 3D data  (details of importing the BNS simulation data into our calculations, using \textsc{pluto}, are described in Appendix~\ref{appendix}).
The original cylindrical average of the output is imported and re-gridded into {\textsc{pluto}} (details of the computational grid can be found in Appendix~\ref{appendix sub: computational grid}). 
After the data import, we evolve the matter distribution within the code and, with a user-defined boundary prescription, we insert a jet into the environment through a nozzle of a radius $r_\jet$ at the excision height $z_0$, imposing a reflective boundary elsewhere along $z=z_0$ (and an axis-symmetric boundary along $r=0$).
Our choice of the boundary condition at $z_0$ does not influence the outcome of our simulations since the velocity of our injected outflow is much larger than the velocity of the ejecta. In addition, due to the strong power of the jet, the newly formed cocoon quickly dominates the ejecta structure in the innermost part of the ejecta.

We perform our simulations using the massively parallel multidimensional relativistic magneto-hydrodynamic code {\textsc{pluto}} (v4.4.2)\footnote{This is the latest version available during the making of this paper.} \citep{mignone_pluto_2007}. 
The code uses a finite-volume, shock-capturing scheme that integrates a system of conservation laws. The flow quantities are discretized on a rectangular computational grid enclosed by a boundary. 
We use the special relativistic hydrodynamics module in 2.5D cylindrical coordinates (i.e., a 2D grid in cylindrical geometry in which $v_\phi$ is also integrated and evolved with periodic boundary conditions).
We performed our simulations using a linear reconstruction scheme combined with a second-order Runge-Kutta time stepping. 
We also force the code to reconstruct the 4-velocity vectors at each time step. 
We chose a Taub-Matthews equation of state \citep{mignone_equation_2007}, which accounts for both relativistic and non-relativistic material, i.e., $\gamma = 5/3$ for non-relativistic material and $\gamma = 4/3$ for relativistically moving matter. 

The choice of a 2.5D simulation over 3D has implications for both the jet propagation and the development of hydrodynamic turbulence in the flow.
A direct effect of the dimensionality is the direction of the turbulent cascade. 
While in 3D, the cascade goes from large to small scales, in 2D, the opposite is true, as shown by \citet{kraichnan_inertial_1967}, resulting in a flow organizing itself in large eddies and generating different turbulent structures that affect the mixing of the jet material with the ejecta.
Since a 2.5D simulation lacks a degree of freedom, the jet is forced to propagate, staying on-axis.
This results in a numerical effect called ``plug", which consists of matter stalling over the jet and affecting its expansion.
Conversely, in 3D simulations, this feature is absent, and the jet, while carving its way out, wobbles due to flow instabilities and interaction with the surrounding ejecta.
When 2D and 3D runs are compared, the difference is not too stark \citep{harrison_numerically_2018}.

\begin{figure*}
    \centering
    \includegraphics[width=1.\linewidth]{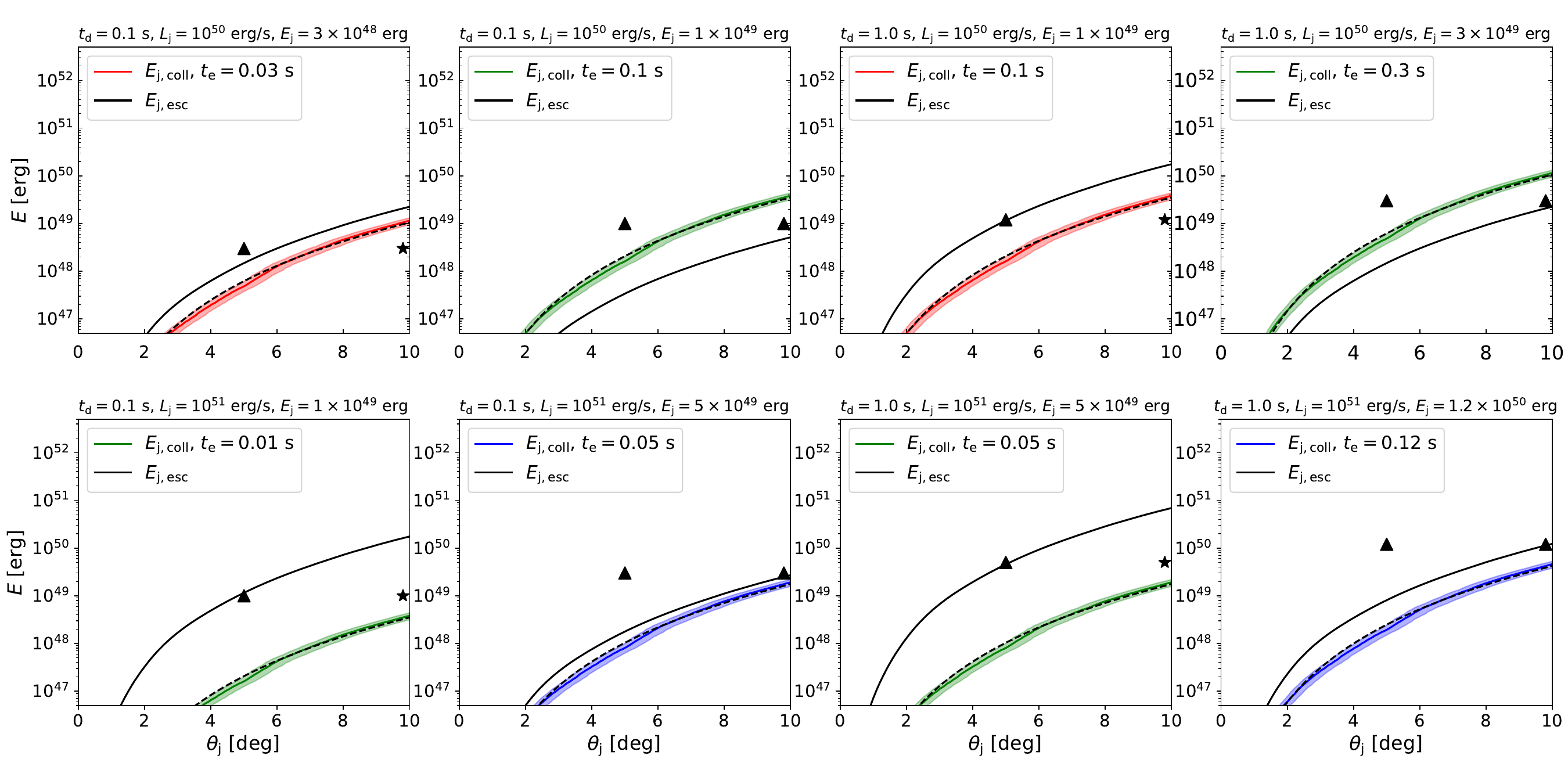}
    \vskip -0.3cm 
    \caption{Jet breakout conditions for injected jets with $\td = 0.1$~s and $\td =1~$s after the merger. \textit{Top}:  $L_\jet = 10^{50}~\erg/\s$; \textit{bottom}:  $L_\jet = 10^{51}~\erg/\s$. The colored lines represent the energy required to collimate the jet to a certain opening angle according to Eq.~\eqref{eq: collimation_energy}. In contrast, the black solid lines represent the energy required for a jet to escape as a function of the opening angle $\theta_\jet$ derived from Eq.~\eqref{eq: escape}. The dots represent the tested cases for $\theta_\jet = 5^\circ$ and $10^\circ$ with triangles (stars) representing successful (choked) jets. The black dashed lines in each panel, superimposed to the colored lines, represent the fit for the collimation curve expressed by Eq.~\eqref{eq: collimation_energy}.}
    \label{fig: collimation_conditions}
\end{figure*}
\begin{figure*}
    \centering \includegraphics[width=0.7\linewidth]{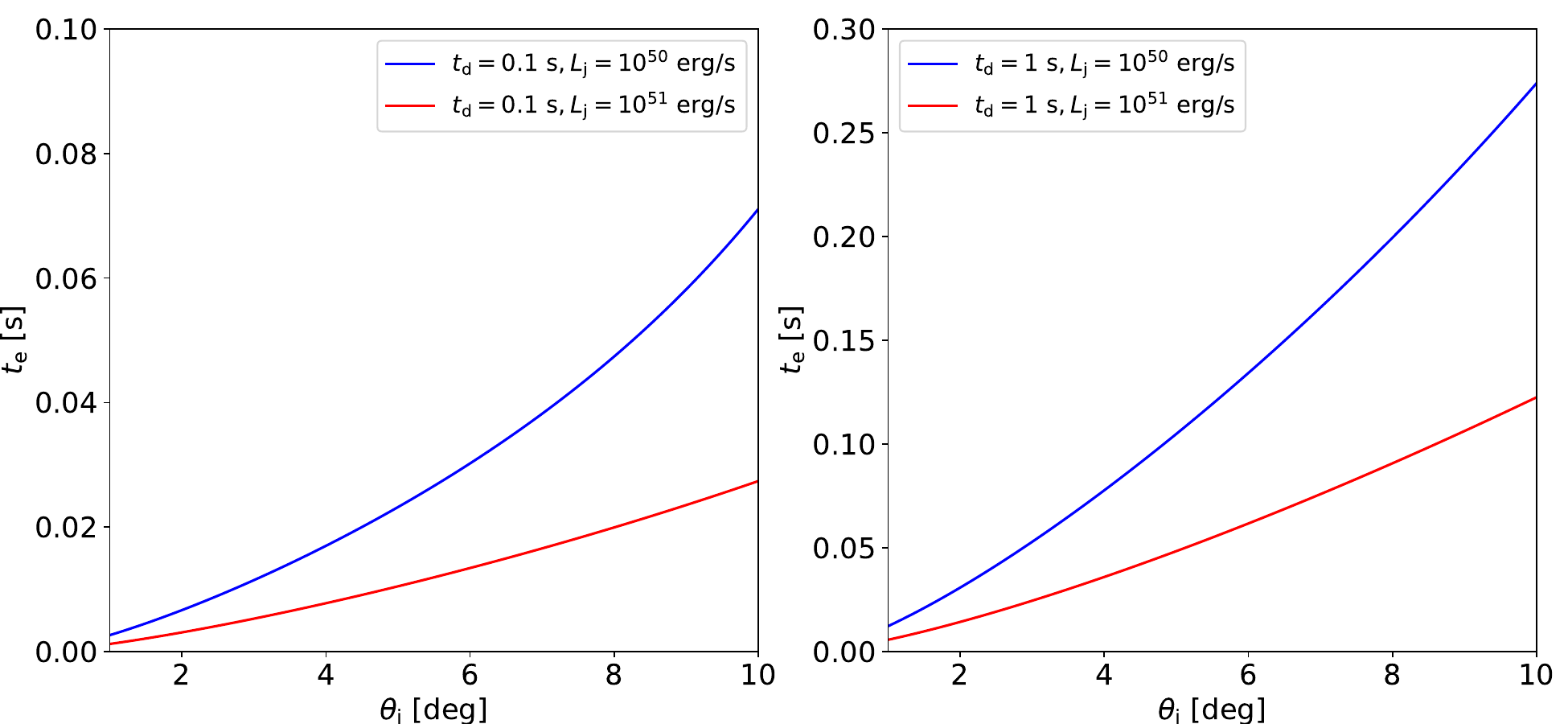}
    \caption{Minimal engine time, $\te$, required for a jet to escape as a function of the injection angle $\theta_\jet$, the luminosity, $L_\jet$, and the time delay, $t_{\rm d}$,  between the merger and jet launch. The lines represent the estimate from Eq.~\eqref{eq: escape2}.
    We can see that in the case of $\td = 0.1~\s$ we get $\te \gtrsim 0.05~\s$, while for $\td= 1~\s$, this value increases to $\te \simeq  0.1 - 0.3~\s$ for jets with $\theta_\jet = 5$--$10^\circ$. 
    All these values for $\te$ are compatible with a sGRB.}
    \label{fig: tengine}
\end{figure*}

\subsection{Jet injection}
\label{subsec: jet_injection}
We inject a jet with constant luminosity $L_\jet$, operating for an engine time $\te$ with an initial bulk Lorentz factor $\Gamma_{0,\jet}$, a density $\rho_\jet $ and a dimensionless specific enthalpy $h_\jet$. 
The hot jet is injected through a nozzle parallel to the $z-$axis with a radius of $r_\jet$ and with an initial opening angle, $\theta_\jet$, which also determines the initial Lorentz factor via $\Gamma_{0,\jet} \simeq 1/(1.4 \theta_\jet)$ (see details on this injection method at \citealt{mizuta_opening_2013} and \citealt{harrison_numerically_2018}). 
We place the nozzle at an initial excision height of $z_0 = r_\jet / \theta_\jet$. 
In all our simulations, we chose $r_\jet = 10~\km$, a size comparable to the Schwarzschild radius of the central BH, allowing sufficient mesh coverage.
The head cross section is given by $\Sigma_\jet = \pi r_\jet^2$. 
This determines $\rho_\jet$ as:
\begin{equation}
    \label{eq: rho_j}
    \rho_\jet  = \dfrac{L_\jet}{\Sigma_\jet h_\jet \Gamma_{0,\jet}^2 \beta_{0,\jet} \mathrm{c}^3} \ ,
\end{equation}
with $\beta_{0,\jet} = \sqrt{1-1/\Gamma_{0,\jet}^2}$.
This choice of the specific dimensionless enthalpy determines the terminal Lorentz factor of the system, i.e. $\Gamma_\infty \simeq h_\jet \Gamma_{0,\jet}$.
The jet's pressure is given by $ P_\jet = h_\mathrm{eff} {\rho_\jet \mathrm{c}^2}$, where $h_\mathrm{eff} =  \left( 5 h_\jet - \sqrt{9 h_\jet^2 + 16}\right)/8$ is derived from Taub-Matthews EOS. We choose $h_\jet = 100$, which gives $h_\mathrm{eff} \simeq 25$. 
For details regarding the computational grid used in {\textsc{pluto}}, we refer to Appendix~\ref{appendix sub: computational grid}. 
At the same time, a more detailed description of the jet injection is found in Appendix~\ref{appendix sub: jet injection}.


\section{Results} \label{sec: results}

\subsection{Collimation criterion}

In \citet{pais_collimation_2023}, we showed how the ambient pressure counterbalances the Poynting flux of a heavily magnetized jet, assuming transverse equilibrium for the magnetic field. 
In our case, the jet is not magnetized, and an equation of state with $\gamma=4/3$ describes the relativistic pressure injected. So, we cannot strictly apply the same criteria for collimation as described in our previous work. 
In particular, if $P_\jet \simeq (h_\jet - 1) \rho_\jet \mathrm{c}^2 / 4$ for hot jets and if $\rho_\jet$ is given by Eq.~\eqref{eq: rho_j}, then the equilibrium condition with ambient pressure reads:
\begin{equation}
\label{eq: collimation}
    P_\ej \simeq \dfrac{L_\jet}{4 \pi r^2 \Gamma^2 \mathrm{c}} \ ,
\end{equation}
where $P_\ej(R,z)$ is the local ejecta pressure, and $\Gamma = \Gamma(R)$ is the Lorentz factor of the flow. 

\begin{figure*}
    \centering
    \includegraphics[width=1\linewidth]{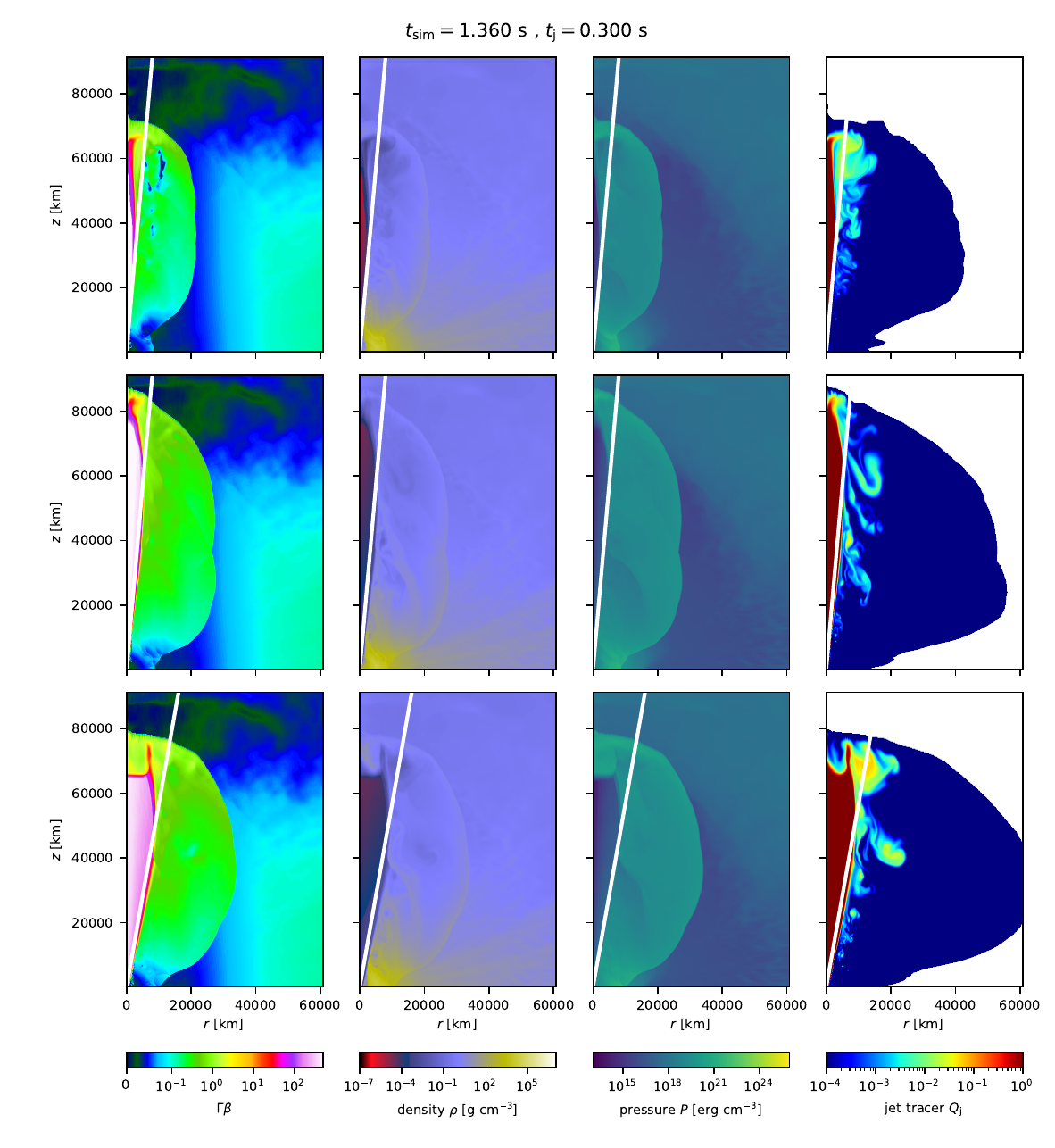}
    \vskip -0.7cm 
    \caption{An example of evolved jets into the BNS-post merger environment for the model $\td = 1~\s$  after $300~$ms of continuous injection for three different cases: \textit{top} - $L = 10^{50}$ erg/s and $\theta = 5^\circ$,  \textit{middle} -  $L = 10^{51}$ erg/s and $\theta = 5^\circ$, \textit{bottom} -  $L = 10^{51}$ erg/s and $\theta_\jet = 10^\circ$. 
    We started the simulation at $t = 1.06~\s$, which is roughly at $\td = 1~\s$ after the collapse of the HMNS. 
    The time $t_\jet = 0.3~\s$ indicates the time since the jet injection has started. 
    The white line represents the opening angle of the jet. From left to right, the panels represent the $\Gamma\beta$ factor, the density $\rho$, the pressure $P$, and the jet tracer (flagged cells with jet material advected with the fluid).}
    \label{fig: system_300ms}
\end{figure*}

For a relativistic jet confined by a non-static medium along the cross-sectional radius $R$ of the jet, the scaling $\Gamma \propto R$ holds \citep{bromberg_hydrodynamic_2007}, which is the same scaling found when transverse equilibrium for the magnetic field is reached for magnetically dominated jets. 
In \citet{pais_collimation_2023}, we derived the collimation curves for the simulation data of \citet{kiuchi_self-consistent_2023}, varying the luminosity of the jet smoothly, determining the corresponding equilibrium contours between the jet pressure and the ambient pressure, and calculating the opening angle of these contours with the assumption of a quasi-conical shape for the jet. 
For the increase in the luminosity, the increase in the jet opening angle was obtained, giving a functional relation between these two quantities. 
Since we use the same background data from the BNS merger as the initial condition for our jet injection, we can fit the collimation curve $L_{\jet,\mathrm{coll}}$ as a function of the angle $\theta_\jet$ with a power law $\simeq L_0 \times \theta_\jet^4$. 
Thus, under the assumption that the luminosity is constant over time and given an angle $\theta_\jet$, the corresponding collimation energy is:
\begin{equation}
\label{eq: collimation_energy}
    E_{\mathrm{coll}} ( \theta_\jet) = \int_0^{\te} L_{\jet,\mathrm{coll}} (\theta_\jet) \de t \approx 3.6 \times 10^{49}~\left( \dfrac{\theta_\jet}{0.1~\rad}\right)^4 \left(\dfrac{\te}{1~\s} \right)~\erg \ .
\end{equation}
We remark that this formula is an approximation, and its prefactor is tailored to the specific energy distribution of the BNS simulation we are using.  Since the luminosity is assumed to be a constant, the collimation energy as a function of the jet angle scales linearly with the engine time.
Analyzing the simulations, we report our fit in Fig.~\ref{fig: collimation_conditions} as a black dashed line over the colored collimation curve lines.

\subsection{Breakout criterion}
\citet{gottlieb_propagation_2022} estimate the breakout criterion for a jet surrounded by anisotropic, fast-moving ejecta.
Their condition for a jet of luminosity $L_\jet$ reads: \begin{equation}
\label{eq: escape}
    E_\jet \equiv L_\jet \times  \te > 150 \left[ \left( \dfrac{\td}{\te} \right)^2 + 2 \right] E_\ej (<\theta_\jet) \times \theta_\jet^2 \ ,
\end{equation}
where $\td$ is the delay time between the merger and the jet launch, $\te$ is the engine time of the jet, $\theta_\jet$ is the jet opening angle, and $E_\ej (<\theta_\jet)$ is the energy of the ejecta within $\theta_\jet$.
We see immediately that if $\td \gg \te$, it becomes energetically difficult for a jet to break out from the ejecta.
If we now express $E_\ej(< \theta)$ as a function of $\theta$, then we fit the data from the BNS simulation as $E_\ej(<\theta_\jet) \simeq E_0 \times  \theta_\jet^k$ (where $k \approx 2)$, and if we express $E_\jet$ as $ L_\jet \times \te $, we get a third-order polynomial in $\te$, which gives us an approximate estimate of the minimal engine time for a jet to break out as a function of $L_\jet$, $\theta_\jet$ and $\td$. 
This estimate, for typical jet parameters, reads:
\begin{equation}
\label{eq: escape2}
    \te \simeq 0.1~\s ~ \left( L_{\jet, 50}^{-1} \theta_{\jet,5}^4   + L_{\jet, 50}^{-1/3} \theta_{\jet,5}^{4/3} t^{2/3}_\mathrm{d,0} \right) \ ,
\end{equation}
where $L_{\jet,50} =L_\jet / (10^{50} \erg/\s)$, $\theta_{\jet,5} = \theta_\jet / 5^\circ$, and $t_\mathrm{d,0} = \td / (1\,\s)$. 
Since $\te \lesssim 1\,\s$ for sGRB jets and the jet is likely to be narrowly collimated, the first term of Eq.~\eqref{eq: escape2} can be neglected. 
This second term agrees with the estimates given by Eq.~(31)  of \citet{nakar_electromagnetic_2019}, even though it was derived under some restrictive conditions that do not hold here. 

We report in Fig.~\ref{fig: tengine} the estimate of the engine time required for the jet to escape from the ejecta using Eq.~\eqref{eq: escape2}.
It is also worthwhile to notice how the prefactor in Eq.~\eqref{eq: escape} quickly increases when $\te \ll \td$, so we will mainly focus on engine times that have a duration between $0.05$ and $0.3$~s. 
We take Eq.~\eqref{eq: escape2} as the minimal estimate for the engine time required by the jets to escape.

Fig.~\ref{fig: collimation_conditions} depicts the escape and collimation conditions for the jet,  based on the simulation analysis of \citet{pais_collimation_2023}. 
The points in the plots show simulated jets: the triangles represent jets that emerged from the ejecta, while the stars represent jets that choked inside the ejecta.   
We stress that these curves \citep[from][]{pais_collimation_2023} represent only estimates of the energy required for the jet to escape and that the actual energy might differ by a factor of a few. 
Moreover, \citet{gottlieb_propagation_2022} develop their analytic theory for a spherical configuration, and here we used an adaptation \citep{pais_collimation_2023} of this estimate to anisotropic ejecta, given by Eq.~\eqref{eq: escape}.
Still, as we show later, the numerical results are consistent with the analytic estimates.

\begin{table}
    \centering
    \begin{tabular}{|c||cc}
        $\td$ [s] & 0.1 & 1\\
        \hline
        $L_\jet$ [erg/s] & $10^{50}$ & $10^{51}$ \\
        \hline
        $\theta_\jet$ [deg] & $5$ & $10$\\
    \end{tabular}
    \caption{ Parameters used in the suite of simulations presented in Sec.~\ref{subsec: simulation}}
    \label{tab: sim parameters}
\end{table}

\subsection{Simulation}
\label{subsec: simulation}
Since the computations are at high resolution and computationally expensive (around 150000 CPU hours), we run the following simulations, changing one parameter each time, as reported in Table~\ref{tab: sim parameters}, giving a total of 8 simulations, 4 for each $\td$. 
To test the collimation effect of the ambient pressure on the jet, we performed two further simulations with a wide opening angle, the first with $\theta_\jet = 35^\circ$, the second with $\theta_\jet = 20^\circ$, and $L_\jet = 10^{51}~\ergsec$, and $\td= 1$ s, for both cases.  
Except for the case study of $L_\jet = 10^{50}~\ergsec$, $\theta_\jet = 5^\circ$ and $\td = 1~\s$ (where we let the injection run for $\te= 550~\millis$), we run our jets for a maximum time of $\te = 300$~ms ($100$~ms) in the case of $\td = 1$ s (0.1 s). 
Although the BH formation happens quite early after the merger in our imported ICs, we explore the case of a late launch at one second to maximize the effect of the ejecta collimation over the jet propagation.
We also run shorter-duration jets for $\td =0.1~\s$ since the ejecta did not expand too far from the center, and the jet is likely to break out earlier.
With a continuous and constant energy injection in our system, we can stop the injection at any time and follow the evolution of the jet-cocoon system. This way, we can determine whether the jet engine time has been sufficient for the jet to break out of the ejecta. 

Furthermore, we are also interested in the jets' evolution until their breakout from the BNS outermost ejecta. 
For this reason, we run several simulations where we stop the injection of the jet at different moments, i.e., $\te = [0.01, 0.03, 0.05, 0.1, 0.12, 0.3]~\s$.  
We performed a lower-resolution re-grid of the simulations (see Appendix \ref{appendix sub: computational grid}) to speed up the jet's evolution after the engine is shut down.

\begin{figure}
    \centering
    \includegraphics[width=1.\linewidth]{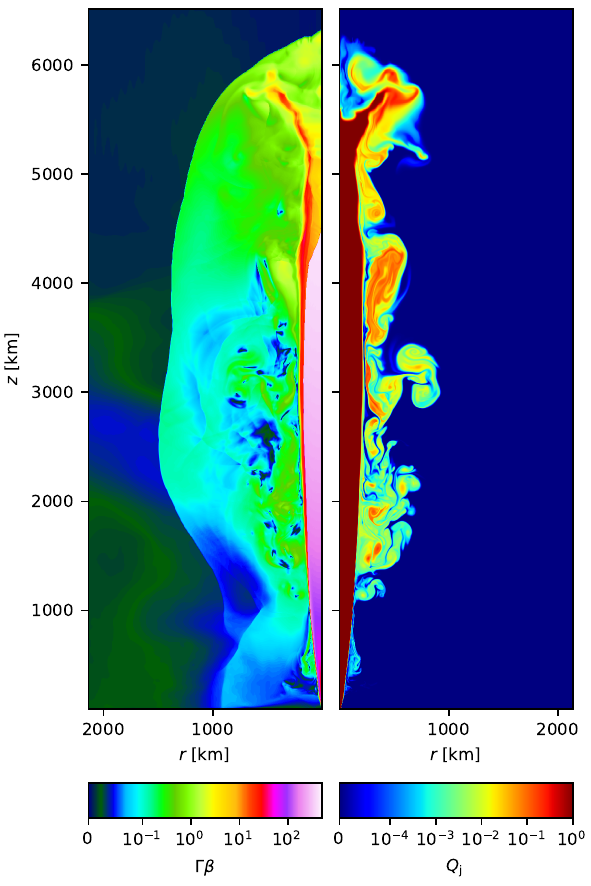}
    \vskip -0.3cm
    \caption{Proper velocity ($\Gamma\beta$) {and jet scalar tracer} maps of a jet initialized with a delay of $\td = 1~\s$, a luminosity of $L_\jet = 10^{51}~\erg/\s$, and an opening angle $\theta_\jet = 5^\circ$ at an early stage of its injection (taken after injecting for $0.025~\s$). At this stage, the jet is still in the ejecta's innermost (and densest) part of the ejecta and drilled through the polar funnel. We can see the shape of the collimation shock, which divides the unshocked jet material from the shocked one. }
    \label{fig: inlet_collimation}
\end{figure}

\begin{figure}
    \centering
    \includegraphics[width=1.\linewidth]{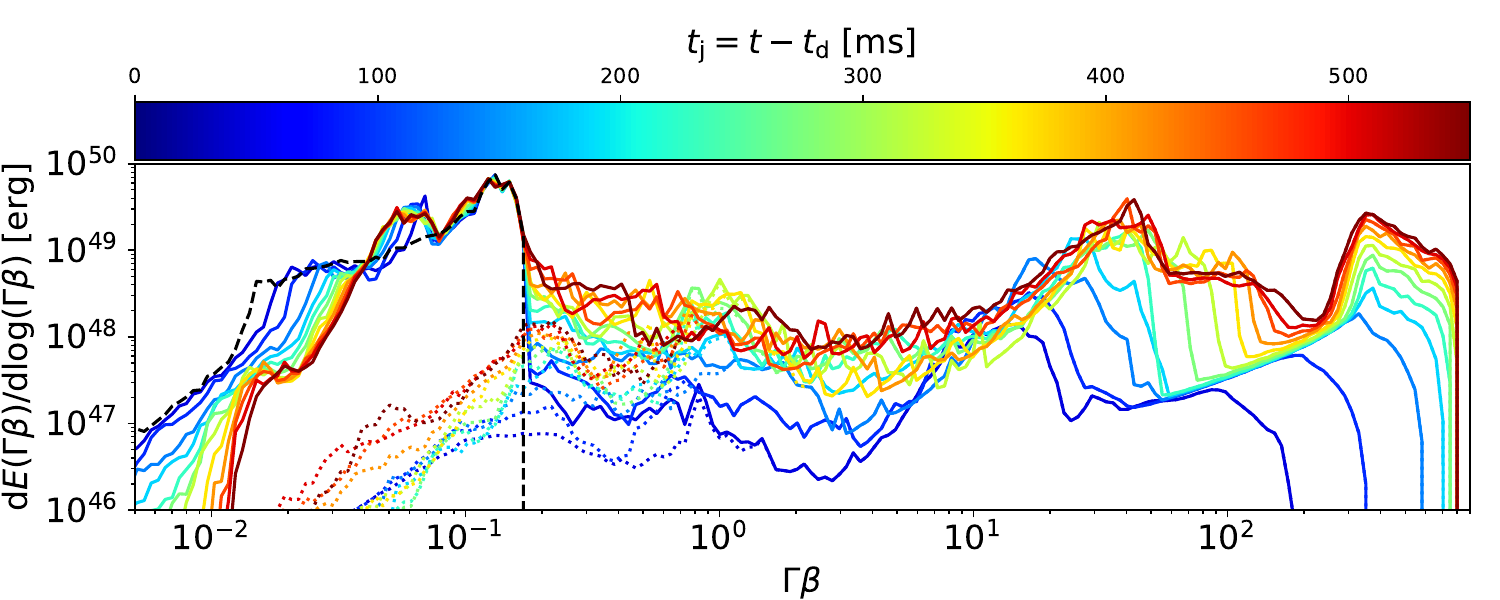}
    \includegraphics[width=1.\linewidth]{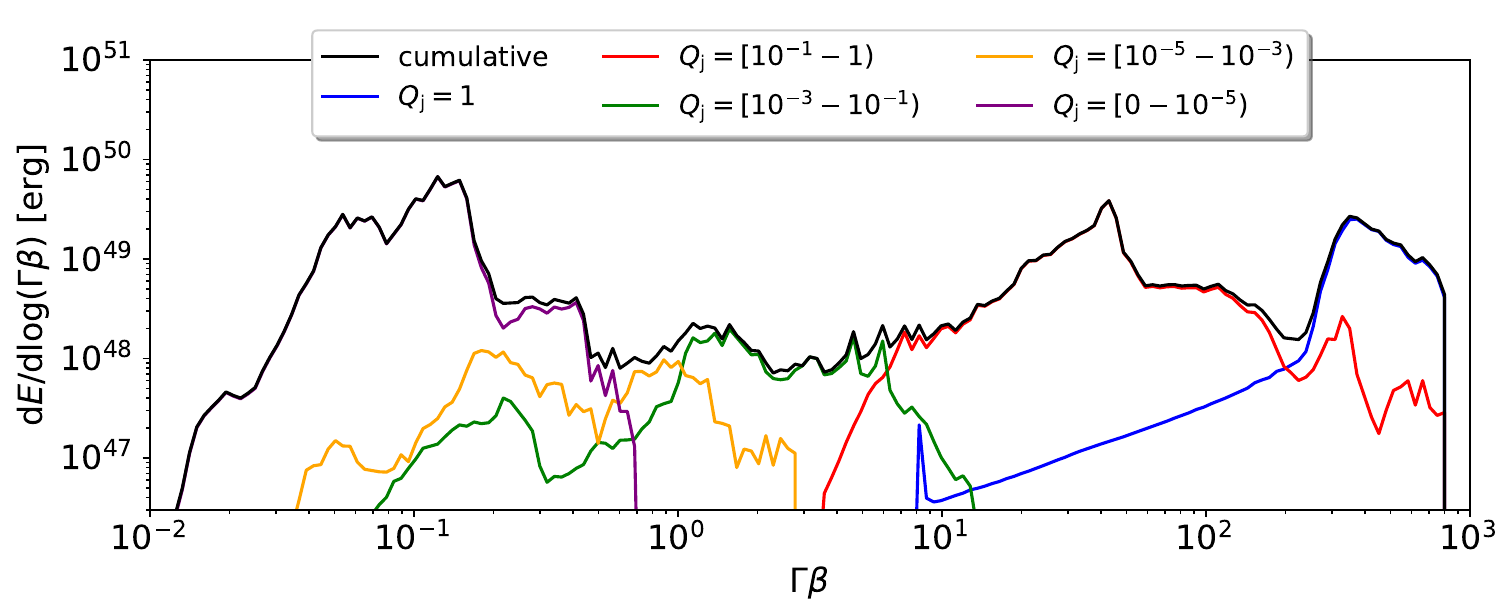}
    \vskip -0.2cm
    \caption{\textit{Top}: Differential energy distribution with respect to the proper velocity $\Gamma \beta $ of a relativistic jet with $L_\jet = 10^{50}~\erg/\s$ and $\theta_\jet = 5^\circ$ injected into the ejecta. The different colors show the jet time evolution during the injection (cold colors for early times, hot colors for late times). The dotted colored lines represent the material associated with the jet. The black-dashed line is the energy of the initial conditions ejecta. Note that we prolonged the simulation just in this case to see the evolution of the distribution till the breakout time $t_\mathrm{bo} \simeq 500~\millis$ after the start of the injection.
    \textit{Bottom}: Different regions of the energy-velocity distribution after $ 550~\millis$ of injection with respect to different thresholds of the jet scalar tracer $Q_\jet$.}
    \label{fig: energy_vs_velocity_1e50t5}
\end{figure}

\subsubsection{Jet evolution for $\td=1~\s$} 
\label{sec: jets} 
In Fig.~\ref{fig: system_300ms}, we show jet simulations with $\td = 1~\s$ and $\te=300~{\rm ms}$. Note that $\td = 1~\s$ is approximately the time delay that took place in  170817A \citep{kasliwal_illuminating_2017, gottlieb_cocoon_2018}.  
The image shows three cases: one with a luminosity of $L_\jet=10^{50}~\erg /\s$ and $\theta_\jet = 5^\circ \sim 0.1~\rad$ (first row), one with the same opening angle but a luminosity of $L_\jet=10^{51}~\erg /\s$ (second row),  one with $L_\jet=10^{51}~\erg /\s$ and $\theta_\jet = 10^\circ \sim 0.2~\rad$ (third row). 
The flow is injected at $z_0 = 100~\km$ and $z_0 = 50~\km$ for $\theta_\jet=5^\circ$ and $10^\circ$, respectively. 
From those, the combination of $L_\jet=10^{50}~\erg$ and $\theta_\jet = 5^\circ \sim 0.1~\rad$ is very similar to the jet properties inferred in GRB 170817A from the afterglow observations \citep[see e.g.][for a summary of the afterglow analysis]{nakar_afterglow_2020}. 
Other simulations, not shown here, are qualitatively similar in their behavior. 
Here and in the subsequent figures, we show the proper velocity $\Gamma\beta$, the density $\rho$, the pressure $P$, and the jet tracer $Q_\jet$, which is a scalar marker that flags the jet material fraction and is passively advected with the fluid, explicitly: $\di_t (\rho Q_\jet)+\nabla\cdot(\rho Q_\jet \bm{v}) = 0$, where $\di_t$ denotes the partial derivative with respect to time.

The jet starts to carve out of the ejecta, filling in $\lesssim 30~$ms the low density-low pressure polar cavities above the compact object. 
As stated in \citet{pais_collimation_2023}, this is a crucial step to let the jet escape: the anisotropic distribution of the ejecta, dense at the equator and dilute at the poles, allows the jetted material to quickly escape the central densest ejecta near the BH ( $\simeq 10^4~\km$ ), entering in a less dense external region, where it is easier for the jet to drill its way out up to the ejecta outer radius $R_\mathrm{ej}$.
As it propagates, the jet forms a cocoon, resulting from its interaction with the merger ejecta. The cocoon grows in size as the jet proceeds its vertical evolution. 
A large proper velocity characterizes the unshocked jet region. 
It resides in the low-density, low-pressure axial cavity and spans an angle roughly the size of the original opening angle. 
The white oblique lines represent the initial opening angle of the expanding jets.
These lines roughly coincide with the opening angle covered by the highest proper velocity material close to the jet axis.
Inspection of the early phases of the jet propagation (see  Fig.~\ref{fig: inlet_collimation}) reveals a collimation shock.
The ambient pressure quickly collimates the jet (in a few tens of milliseconds). 
The collimated jet maintains roughly its original opening angle. 

\begin{figure}
    \centering
    \includegraphics[width=0.99\linewidth]{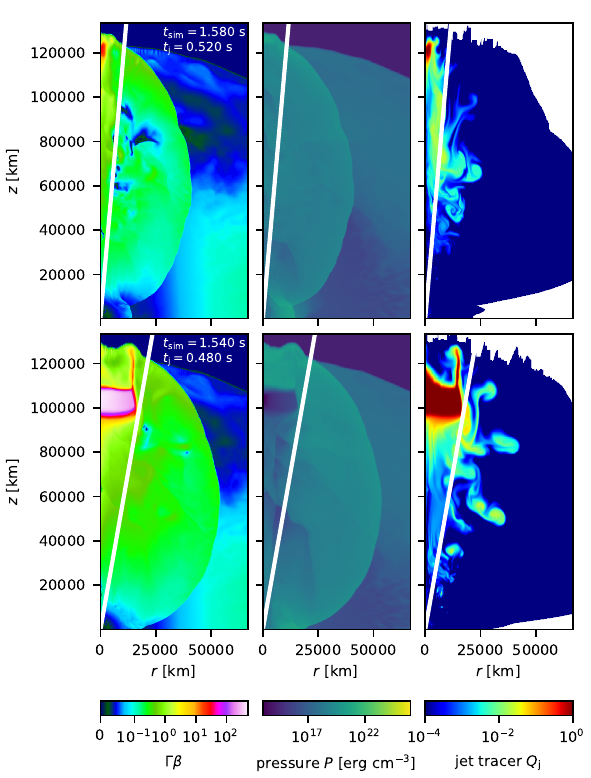}
    \vskip -0.2cm
    \caption{Relativistic jets after we switched off the jet engine running for $\te = 100~\millis$ (top) and $\te = 120~\millis$ (bottom) and let the jet evolve until it reached $R_\ej$ (post-injection phase) for  $L_\jet = 10^{50}~\erg/\s$  and $\theta_\jet=5^\circ$ (top row), and $L = 10^{51}$ erg/s and $\theta_\jet = 10^\circ$ (bottom row). The jet's high-velocity material reaches the ejecta's outer boundary after the jet engine shuts down, leaving jetted material that mixes with the BNS ejecta matter (rightmost panels). In both cases, the breakout happens at $t_\jet \simeq 0.5~\s$.}
    \label{fig: system_post_injection}
\end{figure}

In the top panel of Fig.~\ref{fig: energy_vs_velocity_1e50t5}, we show the time evolution of the differential energy-proper velocity distribution of the jet for the case $L_\jet = 10^{50}~\erg/\s$ and $\theta_\jet = 5^\circ$. 
The hot injected jet accelerates quickly, and a high-velocity energetic tail forms within a few tens of ms.
The energy distribution changes very little after a few hundreds of ms. 
In the case shown, the terminal Lorentz factor equals $\Gamma_\infty = \Gamma_{\jet, 0} h_\jet \simeq 800$. We compare the high-velocity tail to the original ejecta energy distribution (black dashed line in the figure), which has a cutoff around $\Gamma\beta = 0.2$. 
Analyzing the jet energy-velocity distribution as a function of the jet scalar $Q_\jet$ we can associate each region of the curve to a different structure in the jet-cocoon system (bottom panel of Fig.~\ref{fig: energy_vs_velocity_1e50t5}). 
We identify three main peaks in this distribution: a first peak at low velocities, associated with the original ejecta (purple line), a second peak at around $\Gamma\beta \simeq 40$, associated with the jet head and shocked jet material (red line, i.e., compare with the $\Gamma\beta$ maps in Fig.~\ref{fig: system_300ms}), a third peak at $\Gamma\beta \simeq 400$, associated with unshocked jet material (blue line). 
Between the first (low velocity) and second peaks, the energy distribution shows a long plateau associated mostly with the turbulent region of the high-pressure cocoon surrounding the jet (yellow and green lines).

After an injection time of $\te = 0.3~\s$,  the jet and the cocoon enshrining it are still inside the ejecta in all four cases with $\theta = 5^\circ - 10^\circ$ and $L_\jet = 10^{50} - 10^{51}~\ergsec$; however, it has already reached the outermost and shallowest region of the ejecta, which is expanding at $\sim 0.1~$c. If we extend the simulation further, continuing our injection, at around $\te = 0.5~\s$, the jet breaks out of the ejecta. However, the jet engine can be switched off earlier, and the jet can still break out.
As we switch off the engine, the last jet material (the jet \emph{tail}) is launched and moves close to the speed of light in the low-density, low-pressure interior of the jet, trying to catch up with the head, which is causally disconnected and continues to propagate as if the engine were still active. 
When the tail reaches the head, the jet is choked. 
A good marker of jet choking is the disappearance of the central region with the high proper velocity (i.e., $\Gamma\beta \simeq \Gamma_\infty$) and the sideways spreading of the material associated with the jet, which starts to dilate in vortexes. 
If the jet head continues to be causally disconnected from the tail while reaching the end of the ejecta (located, for $\td=1~\s$, at around $z= 120,000~\km$), the jet is unchoked, and it escapes at a high proper velocity along the axis. 

Fig.~\ref{fig: system_post_injection} shows the evolution of two of the simulations shown in the first and second rows after $100~$ms and $120~$ms of injection, respectively. 
In both cases, we see a jet breakout from the outer ejecta. 
This was largely expected from the results of Eq.~\eqref{eq: escape2}. 
To carry out a more detailed investigation, we stopped the jets of also the other simulations initialized with $\td = 1~\s$ earlier than an injection time of $300~$ms to see whether the results of Eq.~\eqref{eq: escape2} are consistent. 

We stopped the jets earlier according to the selected set of parameters ($L_\jet$, $\theta_\jet$) shown in Table~\ref{tab: sim parameters}
and reported our results in the third and fourth columns of Fig.~\ref{fig: collimation_conditions}. 
Each panel shows a case with the same configuration parameters ($L_\jet$, $\te$) and two different values of $\theta_\jet$ ($5^\circ, 10^\circ)$. In the third column, we select the minimal engine time such that the jet with the narrower angle escapes; in the adjacent fourth column, we select the minimal engine time such that the wider jet escapes. 
The triangles indicate jets that could break out from the ejecta, while the stars indicate jets choked inside the ejecta. 
We can see that all points match the expected results for the jet breakout pretty well.

\begin{figure}
    \centering
    \includegraphics[width=1\linewidth]{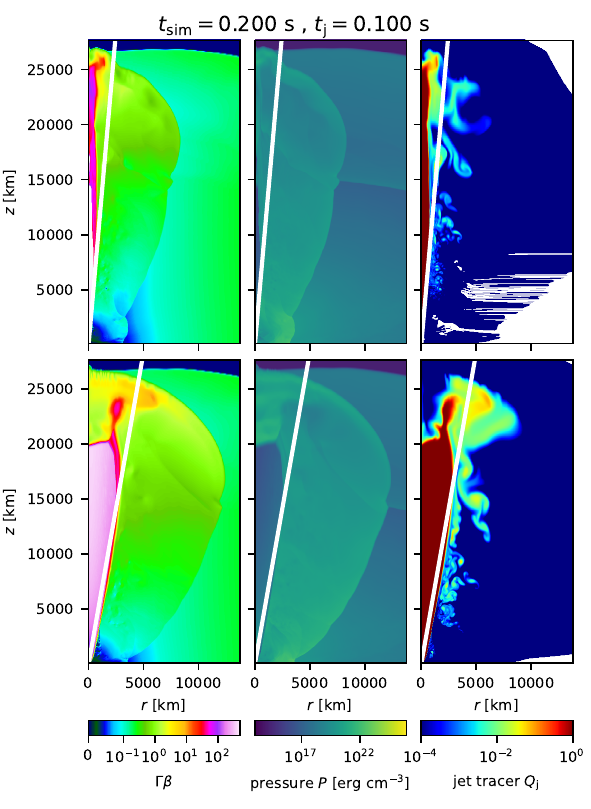}
        \vskip -0.3cm
    \caption{Jet injection for the model $\td = 0.1~\s$ at $t_\engine = 100~$ms after the injection (end of injection) for two cases. \textit{Top}: $L = 10^{50}$ erg/s and $\theta = 5^\circ$ and \textit{bottom}: $L = 10^{51}$ erg/s and $\theta_\jet = 10^\circ$. White lines and plotted physical quantities are the same as the previous multi-panel figures depicting simulation snapshots. For space reasons, we omitted plotting the density map. The top panels correspond to the top row, second column, of Fig.~\ref{fig: collimation_conditions}.}
    \label{fig: system_100ms}
\end{figure}

\begin{figure}
    \centering    
    \includegraphics[width=1\linewidth]{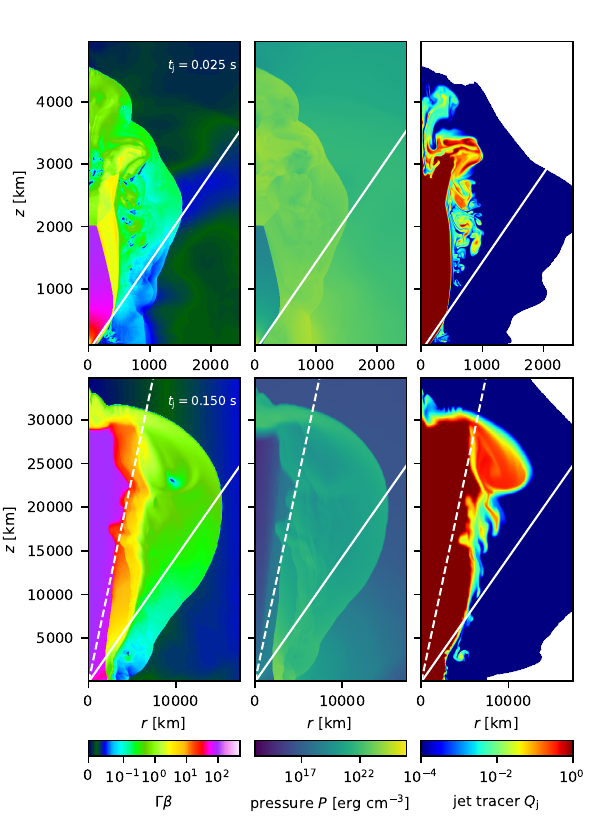}
        \vskip -0.4cm
    \caption{Jet injection with initial conditions $\td = 1~\s$ for a jet with a wide initial opening angle of $\theta_\jet = 35^\circ$ and $L_\jet = 10^{51}$erg/s. The jet is shown in two different moments: (top row) at $t_\jet= 25~\millis$ and (bottom row) at $t_\jet = 150~\millis$. The solid white line is the initial opening angle at which the jet is injected, and the dashed line in the bottom row shows the actual - approximately - opening angle of the jet $\simeq 10^\circ$. The external pressure could collimate the flow to a narrower angle in a few tens of milliseconds. Again, the four panels, from left to right, represent the $\Gamma\beta$ factor, the density $\rho$, the pressure $P$, and the jet tracer (flagged cells with jet material advected with the fluid) }
    \label{fig: 35_deg}
\end{figure}

\subsubsection{Jet evolution for $\td = 0.1~$s}
We now explore the case where the jet is injected at $\td=0.1~\s$ after the merger.
This allows us to study the jet breakout much earlier (which requires less injection time to burst out) but, conversely, from a more dense environment, which already shows the presence of the polar funnels from the compact object. 
We run again four cases with the same luminosities and opening angles as those with  $\td = 1~\s$ (as described in Table~\ref{tab: sim parameters}). 
Fig.~\ref{fig: system_100ms} shows the status of the jets injected at $\td = 0.1~\s$ after 100~ms of injection.
In these cases, we also see that the jets drilled their way out of the ejecta, maintaining their opening angle. 
As predicted by Eq.~\eqref{eq: escape2}, a much shorter engine time is required for the jet to escape since the outer shock of the ejecta $R_{\mathrm{ej}}$ is located much closer to the center at $\simeq 21,000~\km$.
Echoing the $\td=1$ simulations, we decided to test if the engine time required $\te$ to break out from the ejecta is consistent with Eq.~\eqref{eq: escape2}. 
A very short burst ($\te < 0.1~\s$) is sufficient in these cases. We stopped the injection time in all four cases at $\te = [0.01, 0.03, 0.05, 0.1]~\s$. The results are reported in the first and second columns of Fig.~\ref{fig: collimation_conditions}. Again, we express the choked jets as starred dots and the escaping ones as triangles in the plot. Also, in these cases, we verify that the predicted engine times are sufficient to break out from the ejecta. 

\begin{figure}
    \centering
    \includegraphics[width=1\linewidth]{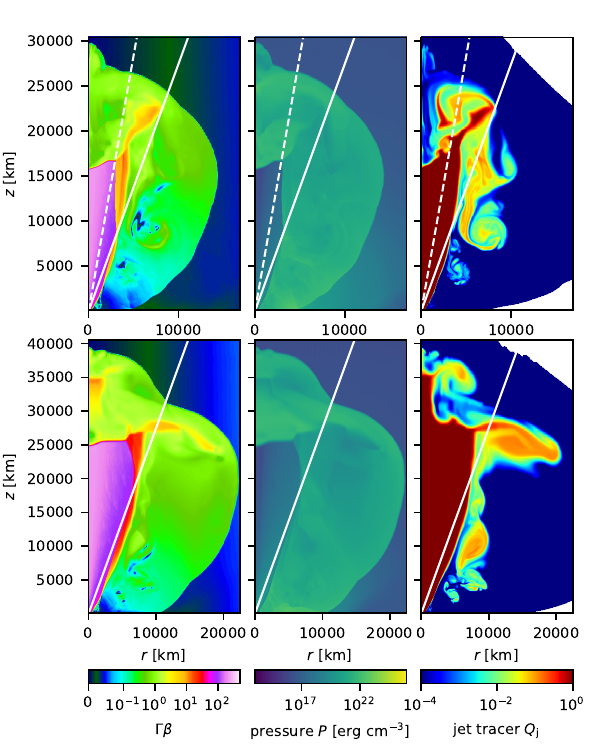}
        \vskip -0.3cm
    \caption{Jet injection with initial conditions $\td = 1~\s$ for a jet with a wide initial opening angle of $\theta_\jet = 20^\circ$ and $L_\jet = 10^{51}$erg/s (top row), and $L_\jet = 5.4 \times 10^{51}~\ergsec$ (bottom row), at $t_\jet = 150~\millis$.}
    \label{fig: 20_deg_comparison}
\end{figure}

\subsection{Initially wide jets}
As a last step in our analysis, we address the collimation effect of the ambient medium on a jet initialized with a very wide angle, i.e., $\theta_\jet = 35^\circ$, and a luminosity of $L_\jet = 10^{51}~$erg/s.
The results are reported in Fig.~\ref{fig: 35_deg}. 
We plotted two different moments of the simulation: $25~$ms after injection and  $150~$ms after injection. 
The collimating effect of the surrounding matter is clearly seen now. The ambient pressure of the ejecta can redirect the material's flux going sideways and collimate the jet to a much smaller opening angle in a few tens of milliseconds. A characteristic cusp-shaped collimation shock is seen at $ t=25$ms. As expected, this leads to an initially cylindrical jet in the region past the collimation shock (see also at $t = 25$\, ms). 
These results are consistent with Eq.~\eqref{eq: collimation} and subsequent Fig.~\ref{fig: collimation_conditions} from which we see that the luminosity of the jet is not sufficient to maintain a wide opening angle; thus, the resulting jet will be collimated to a smaller angle.

The jet later expands sideways to $\sim~10^\circ$, filling the post-merger environment's low density-low pressure polar cavity.
This cavity has an opening angle of $\simeq 10^\circ$, resulting in the jet quickly filling it and propagating with a reduced opening angle, presenting a quasi-cylindrical structure in its evolution (second row).

In Fig.~\ref{fig: 20_deg_comparison}, we show two simulations with an initial jet opening angle of $\theta_\jet = 20^\circ$ with two different luminosities: $10^{51}~\ergsec$ and $5.4 \times 10^{51}~\ergsec$. The latter value is chosen according to Eq.~\eqref{eq: collimation_energy} for the collimation energy for the selected opening angle. The first case presents again a wide jet collimated to a smaller angle by the ejecta pressure; the second is a more energetic jet that expands almost conically at later times. 
 
This has a nontrivial implication: the only way to have very wide jets in this post-BNS merger environment is to inject a large amount of energy in a very short amount of time, i.e., the jet would require a high luminosity ($>10^{52}~\erg/\s$) to achieve a wide opening angle and counteract the presence of the strong ambient pressure.
Thus, it is reasonable to suppose that the jet coming out from the post-BNS merger is relatively narrow ($\theta_{\jet} \leq 10^\circ$) and in line with the observations of the afterglow of GW 170817 \citep{nakar_afterglow_2020}. 

\subsection{The angular distribution}

The angular distribution of the total energy of the outflow (the jet, the cocoon, and the ejecta) has significant implications for the EM signals: the prompt gamma rays that arose from the cocoon in the case of GRB 170817A, the observed afterglow that arose first from the cocoon and then from the jet, and a possible late radio afterglow arising from the rest of the ejecta. 

\begin{figure}
    \centering
    \includegraphics[width=0.99\linewidth]{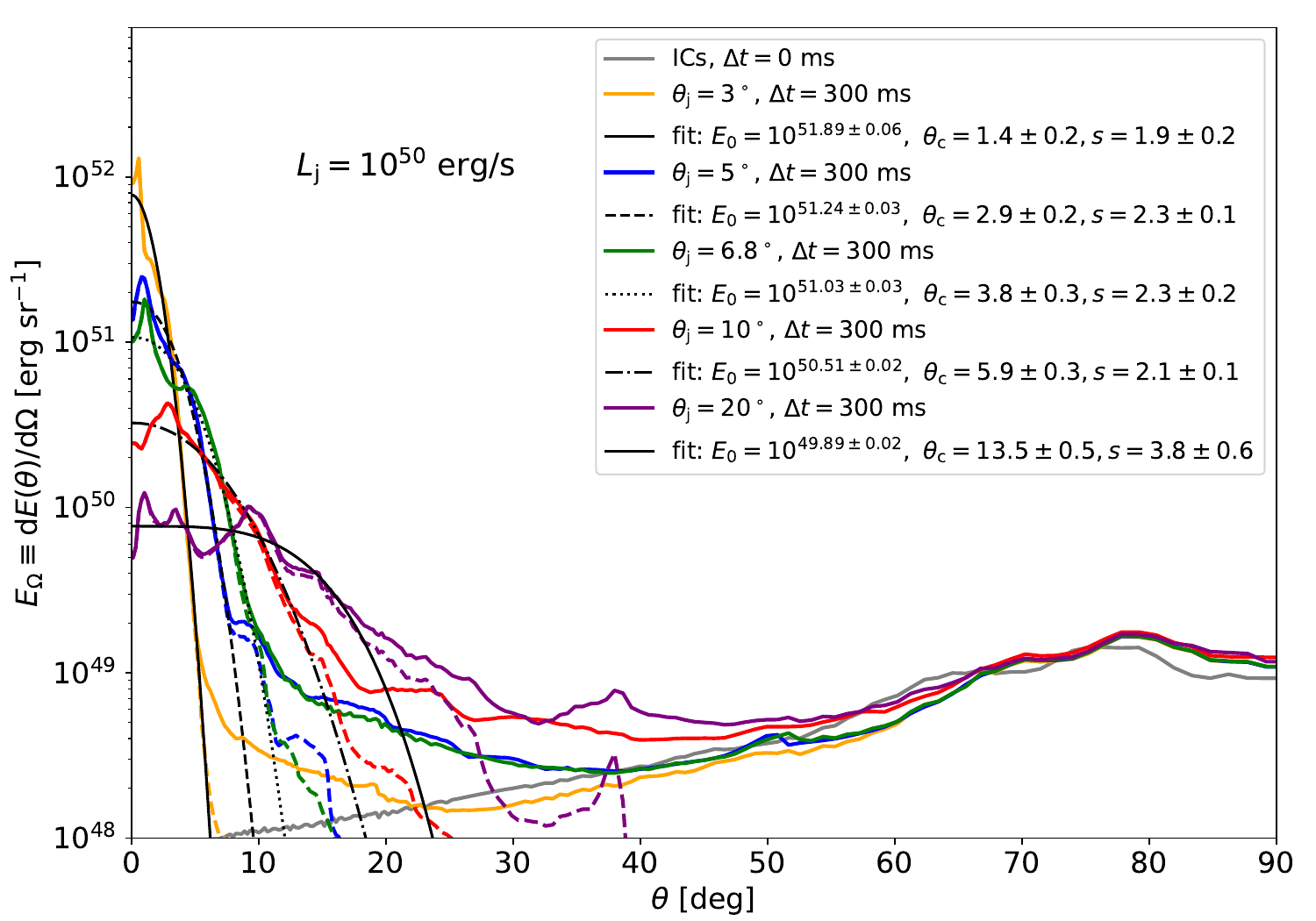}
    \includegraphics[width=0.99\linewidth]{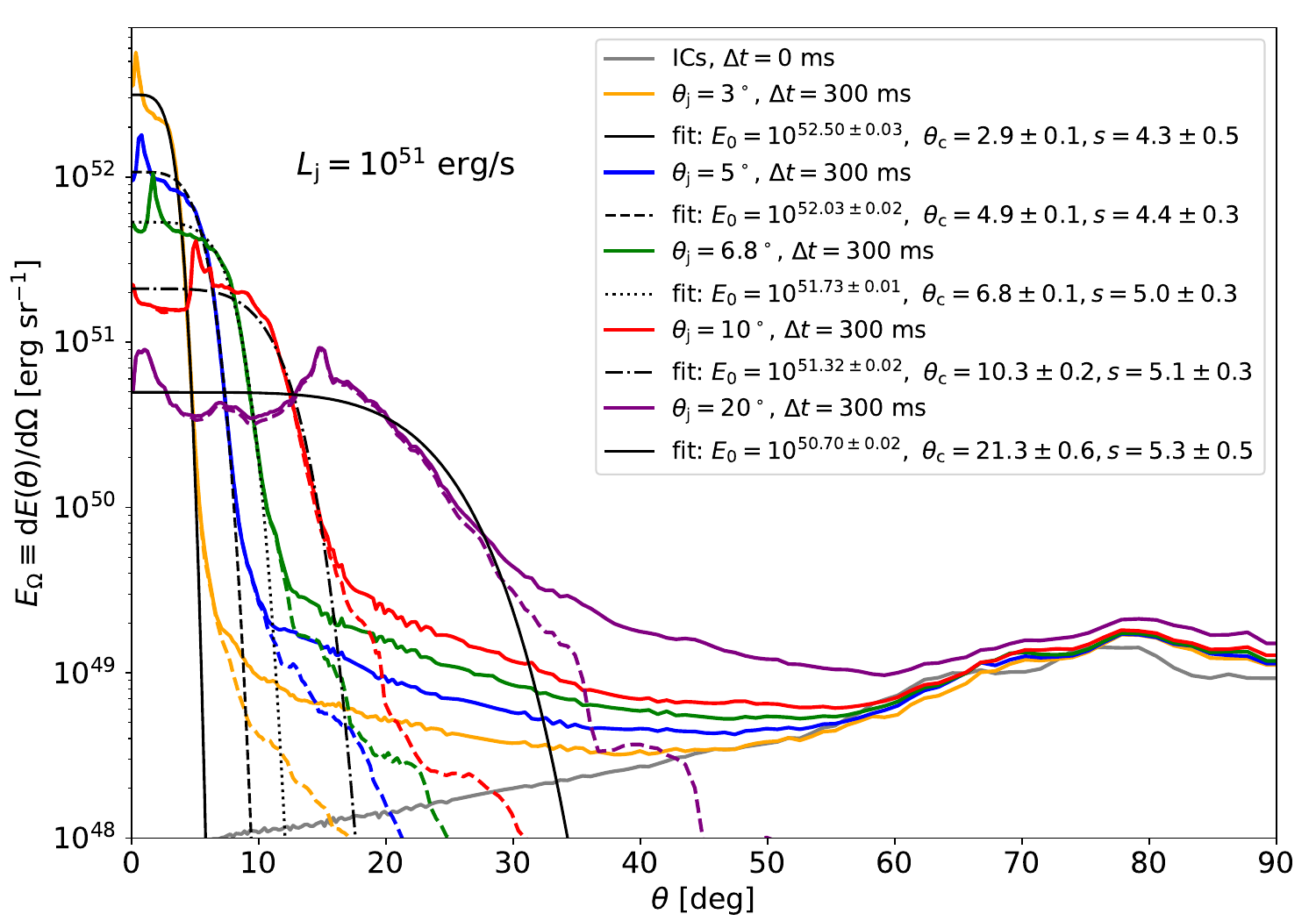}
    \caption{
    Jet energy per solid angle for jets with constant luminosities of $L_\jet = 10^{50}~\erg/\s $ (top panel) and $10^{51}~\erg/\s $ (bottom panel). In each figure, we report the distributions for $\theta_\jet = 3^\circ $ (orange line), $\theta_\jet = 5^\circ $ (blue line), $6.8^\circ$ (green line), $10^\circ$ (red line), and $20^\circ$ (purple) with an injection time of $\te=300~\millis$, and their respective generalized Gaussian fits (solid, dashed, dotted, dashed-dotted, solid black lines). The dashed colored lines for each run indicate the energy per solid angle of the sole jet material. The fit parameters refer to Eq. \eqref{eq: dEdOmega}. The grey line represents the initial ejecta's energy per solid angle. The minima in the calculated distribution are most likely due to the artificial axial symmetry that produces a ``plug" ahead of the jet. The total energy for each jet model is $3\times 10^{50}~\erg)$.}
    \label{fig: energy_vs_angle1}
\end{figure}

Fig.~\ref{fig: energy_vs_angle1} depicts the differential isotropic equivalent energy distribution for four different runs with $\theta_\jet = [3,5,6.8,10,20]^\circ$, and two luminosities $L_\jet = 10^{50}~\erg/\s$ and $10^{51}~\erg/\s$, taken at $\te = 300~\millis$. For each luminosity, the runs have all the same total energy $E_\jet(\te) = 3 \times 10^{49}~\erg$ and $3 \times 10^{50}~\erg$, respectively. 
As expected, the narrower the opening angle, the more collimated the energy distribution is. For the $10^{50}~\ergsec$ runs, the effect of the collimation by the ejecta is evident since the fit shows a smaller opening angle than the initial one, and the curve looks closer to a standard Gaussian. Overall, we find a core with rather sharp edges surrounded by a lower energy cocoon material (from the jet up to $\sim 50^\circ$--$60^\circ $ and ejecta material from $\sim 50^\circ$--$60^\circ $ to the equatorial plane). 

We fit these curves  with a generalized Gaussian function:
\begin{equation}
\label{eq: dEdOmega}
    E_\Omega(t,\theta) = \dfrac{\de E}{\de \Omega} \equiv \dfrac{1}{2\pi\sin\theta}\dfrac{\de E}{\de \theta } =  E_0 (t) \exp\left[- \dfrac{1}{2} \left(\dfrac{\theta}{\theta_\mathrm{c}(t)} \right)^{s(t)} \right] \ ,
\end{equation}
where $\theta_\mathrm{c}$ is the angular spread of the core of the jet, $E_0(t)$ is proportional to the total energy of the jet at the time $t$, and $s$ is a shape parameter that depends on time. This fit is suitable for the core region that is dominated by the jet and has a rather steep profile. 
We notice that the more the jet evolves, the more $\theta_\mathrm{c}$ saturates to $\theta_\jet$, the shape parameter $s$ stabilizes to a constant value, and $E_0$ increases following the same time evolution of $E_\jet = L_\jet \times t$. 
The jet's core angle $\theta_\mathrm{c}$ is in excellent agreement with the jet's initial opening angle within the error, and the wider the opening angle, the higher the jet's shape parameter is (indicating a flatter distribution within $\theta \leq \theta_\mathrm{c} \simeq \theta_\jet$).
A more in-depth overview of the parameters regulating the jet shape and their correlations can be found in Appendix~\ref{appendix: jet_shape}.

\subsection{Afterglow light curves}

\begin{figure}
    \centering
\includegraphics[width=0.99\linewidth]{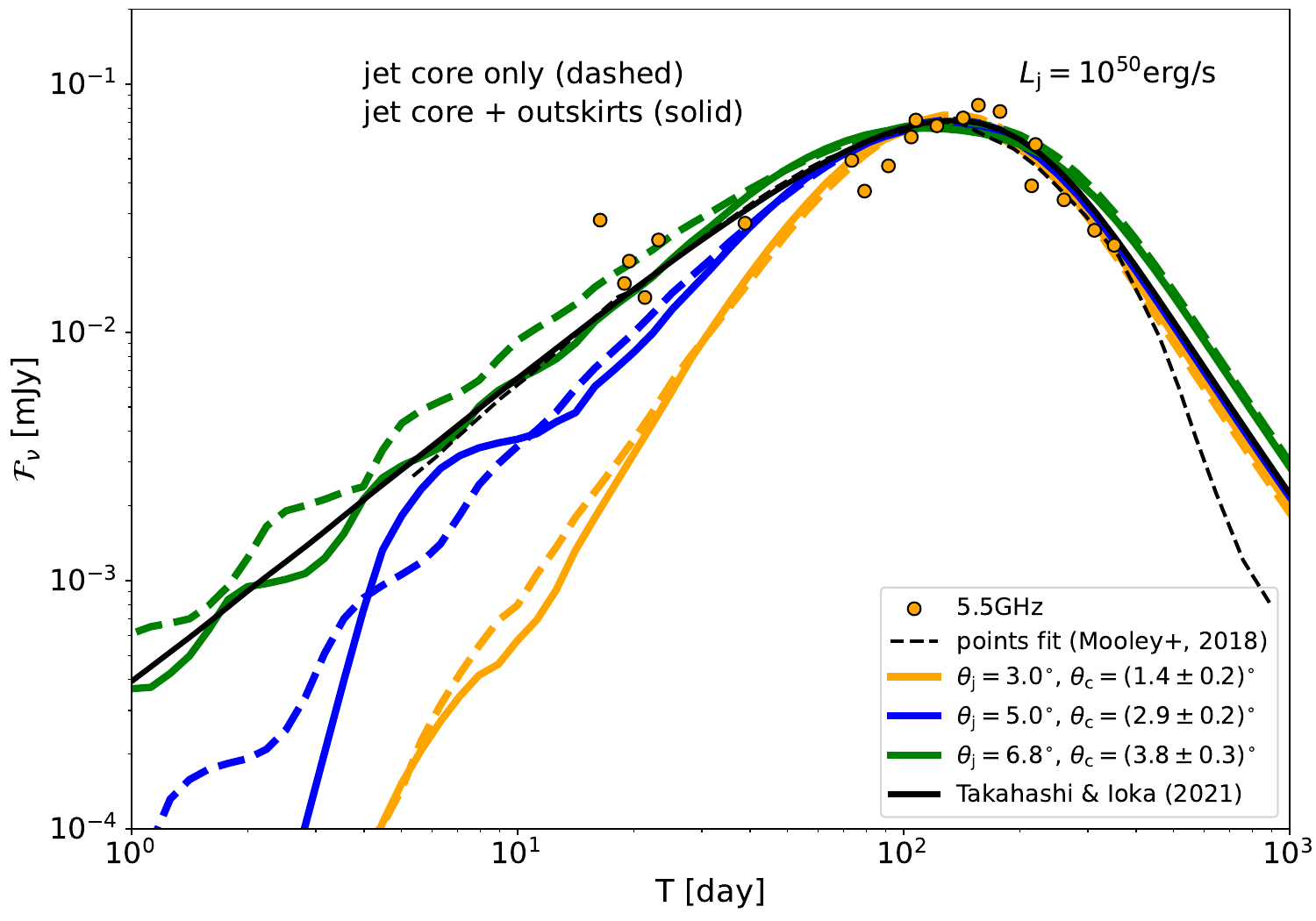}
\includegraphics[width=0.99\linewidth]{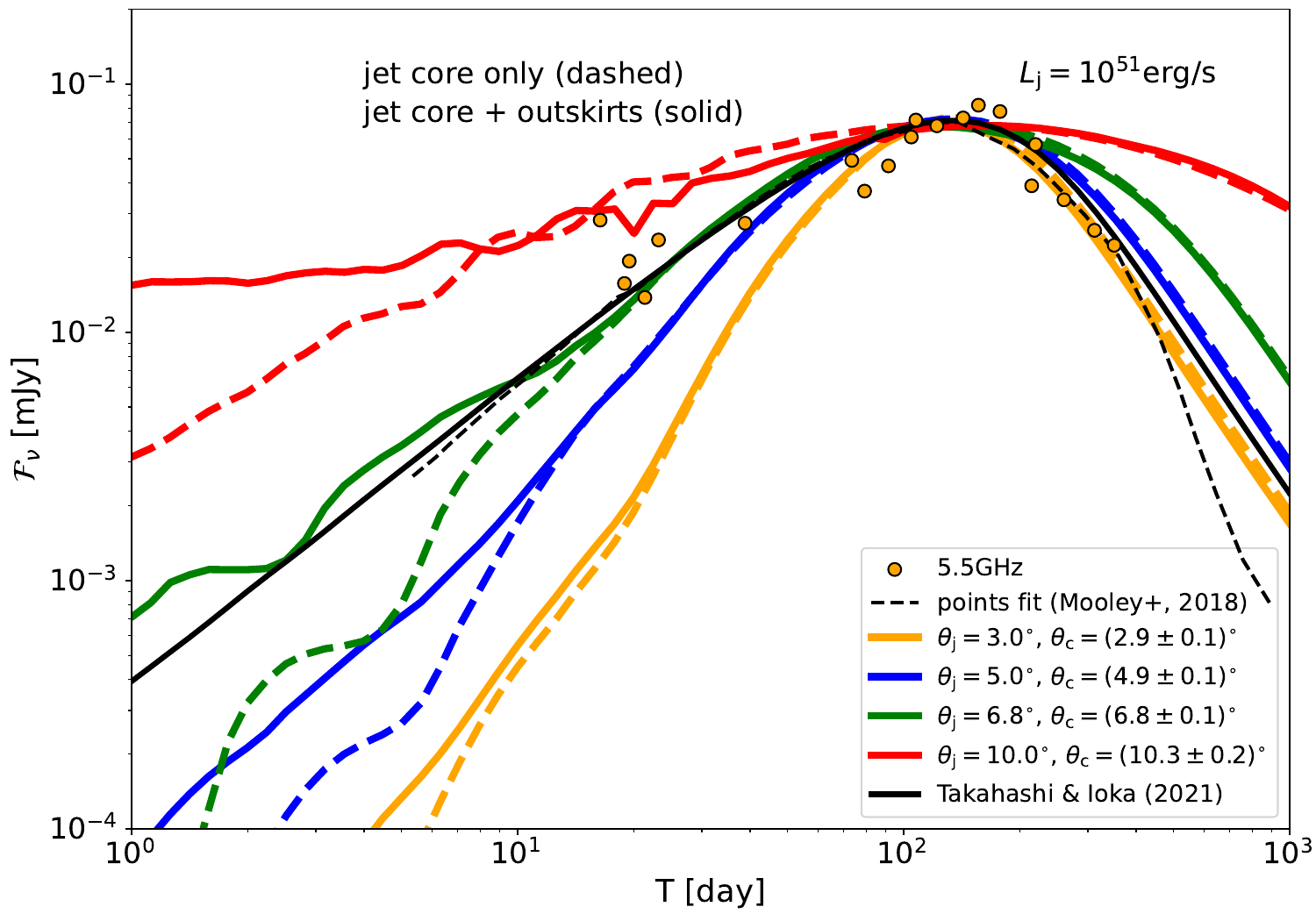}
    \caption{Afterglow light curves at $5.5~\GHz$ for $\theta_\obs = 18^\circ$ of several jet models with different opening angle, $L_\jet = 10^{50}~\ergsec$ (top) and  $L_\jet = 10^{51}~\ergsec$ (bottom), that use the $E_\Omega$ modeled after the fitting curves shown Fig.~\ref{fig: energy_vs_angle1} (same color scheme). On the upper panel, we report only three synthetic light curves out of four since the model for $\theta_\jet = 10^\circ$ with $L_\jet = 10^{50}~\ergsec$ has unrealistic values for $n_0$ and $\epsilonB$. The radio points and relative fit (black dashed line) are taken from \citet{troja_year_2019}. The dashed lines are the afterglow from the generalized Gaussian fits only, while the solid lines are the light curves of the jet, including the energy distribution outside the Gaussian fits. For comparison, alongside the value of the jet opening angle, we report $\theta_\mathrm{c}$ from the fit.}
    \label{fig: afterglow_curves_at_18deg}
\end{figure}

We use the simulation results shown in the previous sections to calculate the light curves of the afterglow emission. We implement a model similar to the one described in \citet{takahashi_inverse_2020} \citep[see also][and references therein]{van_eerten_off-axis_2010}. 
We then compare the result to the actual data points of the afterglow curves of GW170817. For a more detailed description of the methodology implemented to calculate the afterglow emission, including formulas, we redirect the reader to Appendix~\ref{appendix: afterglow model}.

For the afterglow curve, the viewing angle is set at $\theta_\obs = 18^\circ$ as this is approximately the angle found by \citet{hotokezaka_hubble_2019, ghirlanda_compact_2019} who take also into account the VLBI observations of the centroid motion \citep[see also][for a comparison of different estimates of the opening and viewing angles]{nakar_afterglow_2020}.
We fix the spectral index at $p = 2.17$, consistent with models of diffusive shock-accelerated supra-thermal electrons. 
The ISM density $n_0$, $\epsilon_\mathrm{e}$ (fraction of accelerated electrons energy density over the kinetic energy density at the shock), and $\epsilon_\mathrm{B}$ (the fraction of the magnetic energy density over the kinetic energy density at the shock), are linked by a degeneracy relation, which gives the dependency of one of the three parameters of our choice concerning the other two, for a fixed jet model and a fixed $\theta_\obs$. 

\begin{table}
    \centering
    \begin{tabular}{cccccc}
        $\log_{10} L_\jet $ & $\theta_\jet$ [deg] & $\log_{10} E_\jet$ & $\log_{10} n_0 $ &
        $\log_{10}\epsilone$ & $\log_{10} \epsilonB$ \\ \hline
        50 & 3 & 49.48 & -3.78 & -1 & -2.20 \\
        50 & 5 & 49.48 & -4.48 & -1 & -1.41 \\
        50 & 6.8 & 49.48 & -4.87 & -1 & -1.04 \\
        51 & 3 & 50.48 & -3.05 & -1 & -4.25 \\
        51 & 5 & 50.48 & -3.77 & -1 & -3.50 \\
        51 & 6.8 & 50.48 & -4.57 & -1 & -2.75 \\
        51 & 10 & 50.48 & -6.69 & -1 & -0.90 \\
    \end{tabular}
    \caption{Afterglow parameters used in Fig.~\ref{fig: afterglow_curves_at_18deg} to calculate the light curves at $5.5~$GHz. For brevity, we omitted the units for $L_\jet$ (in $\ergsec$), $E_\jet$ (in $\erg$), and $n_0$ (in $\cm^{-3}$).}
    \label{tab: afterglow_parameters}
\end{table}

Fig.~\ref{fig: afterglow_curves_at_18deg} depicts the afterglow curves at $5.5$ GHz for two values of the luminosity $L_\jet = [10^{50}, 10^{51}]~\ergsec$ and four lower resolution jet models with $\theta_\jet = [3,5,6.8,10]^\circ$ and $L_\jet = 10^{51}~\erg/\s$ for a  viewing angle of $\theta_\obs = 18^\circ$.  
The afterglow curves are obtained using the $E_\Omega$ distribution at $\te=300~\millis$ for all models when the jet engine is still active. Thus, the energy deposited by the jet with $L_\jet = 10^{50}~\ergsec$ ($10^{51}~\ergsec$) is $E_\jet = 3 \times 10^{49}~\erg$ ($3 \times 10^{50}~\erg$).
Switching off the injection for the jets at this time will not change their $E_\Omega$ profile too much as long as the jet has already crossed most of the ejecta and breaks out.
The curves are compared to the fits obtained by \citet{mooley_mildly_2018} (black dashed curve) and \citet{takahashi_inverse_2020}, which uses the average afterglow parameters of \citet{troja_year_2019}.

Fig.~\ref{fig: afterglow_curves_at_18deg} depicts both the afterglow light curves obtained considering the full $E_\Omega$ distribution (solid curves) and the generalized Gaussian fit (dashed curves) that considers only the jet material. 
In the latter case, we have excluded the slow material associated with the ejecta, selecting only the fast matter with a jet tracer value $Q_\jet > 10^{-5}$ and $\Gamma\beta \geq 1$. 

The first thing we notice is that, while there was no attempt to guide the calculations to fit a given observed afterglow (apart from setting $n_0$ and $\epsilon_B$ to fit the time of the peak and its magnitude), the model with $\theta_\jet = 6.8^\circ$ for the $10^{50}$\,erg/s jet gives a good fit to the observed light curve.  
The effective opening angle of the jet after collimation is $\theta_{\rm c}=3.8^\circ$. Thus, the ratio of the jet opening angle to the viewing angle, $\theta_\jet/\theta_\obs$, is consistent with the one obtained for synthetic jets estimates  \citep[see][]{nakar_afterglow_2020}. 
The $10^{51}$ erg/s jet is less collimated, and as it is too wide, $\theta_{\rm c}=6.8^\circ$, the resulting afterglow is somewhat inconsistent on the declining part of the afterglow.

The rising portion of the light curves for the jets with $\theta_\jet = 3^\circ$ and $5^\circ$ are below the observed data. Additionally, these jets have a narrow peak. This is due to the sharp edges of the energy distribution that decline faster than a Gaussian. 
For $\theta_\jet = 10^\circ$ and $L= 10^{51}$ erg/s, the peak becomes much too wide and overestimates both the rising and the declining phase of the curve at early and later times. The parameters of these curves are found in Table~\ref{tab: afterglow_parameters}, in which, for simplicity, we keep $\epsilone = 0.1$ constant for each model and we vary $n_0$ and $\epsilonB$ to adjust the position of the peak time $T_\peak$ and the height of the peak flux $\mathcal{F}_{\nu, \peak}$.


\section{Summary and Conclusions} 
\label{sec: conclusions}

We presented a set of 2.5D  special relativistic simulations of unmagnetized relativistic jets launched into a realistic post-BNS merger environment based on the neutrino radiation GR-MHD simulation by \cite{kiuchi_self-consistent_2023} that was evolved for more than 1 second after the merger.
We explored a variety of configurations for relativistic jets launched in proximity of a compact object and expanded in a post-merger environment.
We verified the semi-analytical prediction presented in \cite{pais_collimation_2023} about the jet escape and collimation and validated our results for a jet launched at $0.1$~s and $1$~s after the merger.
We summarise our findings as follows:

\begin{itemize}
\item The low-pressure, low-density polar funnel that arises in the BNS simulations makes it easier for the jet to emerge from the ejecta. 
\item For wide jets, the ambient pressure of the ejecta counterbalances the pressure of the jet while it propagates and collimates it. 
\item We confirm the predictions derived by \cite{gottlieb_propagation_2022} for a minimum engine operative time for the jet to escape.  The numerical simulations verified that the predicted engine time $\te$  allows the jet to escape from the system after the shutdown of the central engine driving it.
\item We explored the conditions of a late jet launch at $\td = 1~\s$ and an early jet launch at $\td=0.1~\s$, finding a significantly shorter engine time was required to escape in case of an early launch at $\td=0.1~\s$, i.e., $0.01-0.05~\s$ vs. $0.1-0.3~\s$ for a launch with a delay of $\td = 1~\s$. 
\item We verified that the choice of the enthalpy (and thus of the terminal Lorentz factor) does not influence our results as long as the jet is hot.
\item We tested the effect of the ejecta thermal pressure in collimating the jet injecting two very wide jets ($\theta_\jet = 20^\circ \simeq 0.35~\rad$ and $35^\circ \simeq 0.6~\rad$) and following their evolution in the ejecta. 
We showed how the dense inner ejecta can collimate the jet to an opening angle of about $10^\circ$ in a few tens of milliseconds while the jet is still located close to the compact object where it is surrounded by dense ejecta matter.
\item  We compared our synthetic light curves generated from the energy profile of our simulations with the observed afterglow data of GW170817 at $5.5~\GHz$. 
Our isotropic equivalent energy distributions are characterized by sharp edges around the jet's opening angle and decline faster than a Gaussian. 
Nevertheless, the resulting synthetic light curves can match radio data for collimated jets with $E_\jet = 3 \times 10^{49}~\ergsec$,  $\theta_\jet = 6.8^\circ$ (corresponding to $\theta_\mathrm{c} = (3.8 \pm 0.3)^\circ$) and a viewing angle of $\theta_\obs = 18^\circ$. In contrast, for smaller (larger) jets' opening angles, the light curve's peak becomes too narrow (wide). 
These values are roughly in agreement with those obtained by fitting jet models to the light curve \citep{hotokezaka_hubble_2019, ghirlanda_compact_2019,lamb_optical_2019,takahashi_inverse_2020}.
The more powerful $E_\jet = 3 \times 10^{50}~\ergsec$  jets are uncollimated (i.e., they roughly maintain their original opening angles). These jets decline or rise too steeply and do not match the radio data well. 
\item These calculations that follow the jet propagation in the ejecta enable us to break the degeneracy between the total energy of the jet, $E_\jet$, and the external density, $n_0$ that exists in ad hoc fittings of jet profiles to the light curve \citep[e.g.][]{nakar_afterglow_2020}.
\end{itemize}

The novelty of this work is that we can study the jet launch much later than previous works due to the 1-second-long evolved initial conditions, which tracked the ejecta's expansion into the circumbinary environment. 
As we have seen, it is more difficult for the jet to break out from the ejecta in case of a late launch, thus putting stronger constraints on the jet engine. 
This is important, at least in the context of modeling GW 170817A, as it has been suggested that there was about a 1-sec delay between the merger and the jet launch \citep{kasliwal_illuminating_2017,gottlieb_cocoon_2018,nakar__2018} in that event. 

Because of this long duration, there is an external decelerating shock wave propagating outwards at a velocity of $0.1~$c in the artificially imposed ``numerical vacuum"  that has a floor value for the density of $\approx 0.17~\g/\cm^3$. This does not influence our main concerns here: jet collimation and breakout. However, it influences the fine details of the shock breakout from the ejecta that is needed in case one wishes to calculate the signature of the cocoon breakout - which was, most likely, the cause of the observed $\gamma$-rays in GRB 170817A \citep{kasliwal_illuminating_2017,gottlieb_cocoon_2018,nakar__2018}.

Modeling the observed sGRB detected in GRB 170817A would require following up the shocked cocoon until it catches up with the low mass very-high velocity component of the ejecta so that the optical depth of the shock is of order $\tau \simeq 1$. This would require a longer run of our simulations, which would take at least 2 seconds. However, for a realistic simulation, the external ``numerical vacuum" in such a case should be much lower so that it does not slow down the fast component of the ejecta. 
Namely,  the circumbinary atmosphere surrounding the post-merger ejecta has to have a mass smaller than the fast-moving ejecta (i.e., $ M_\mathrm{atmosphere} < 10^{-4} \msun$) within the simulation box of about $10^{11}-10^{12}$ cm. We leave this analysis to future work.  Additional future work would involve exploring the propagation of a Poynting flux-dominated jet in this ejecta.

In this work, we have demonstrated that for jet luminosities of $10^{50}$--$10^{51}$\,erg/s an engine time $\te$ of $0.1$--$0.3~\s$ are required for a late launch ($\td = 1~\s$) and $\te \sim 0.01$--$0.05~\s$ in the case of an early launch with a time delay of $\td = 0.1~\s$.
Recalling that the jet in GW 170817 was less powerful compared to typical sGRBs, we note that a longer duration would be required for less powerful jets to break out. 
The jet coming out from the system is narrowly collimated.
The collimation is supported by the narrow funnel in the initial outflow configuration and does not exceed an opening of $\theta_\jet \lesssim 10^\circ$. A wide jet injected into the ejecta is quickly collimated to a smaller opening angle.


\section*{Acknowledgements}
We thank the anonymous referee for their constructive report, which helped improve this work's quality.
MP thanks Riccardo Ciolfi for his valuable comments. 
This work was supported by the ERC grants Multijets and by the Simons Collaboration on Extreme Electrodynamics of Compact Sources (SCEECS) (TP, MP), and by Grant-in-Aid for Scientific Research (Grant Nos.~JP20H00158 and JP23H04900) of Japanese MEXT/JSPS (MS), and by Grant-in-Aid for Scientific Research (Grant Nos.~JP23K25869, and JP23H01172) of Japanese MEXT/JSPS (KK). MP acknowledges additional support from the European Union under NextGenerationEU via the PRIN 2022 Project "EMERGE" Prot. n. 2022KX2Z3B. 
This work used the computational resources of the supercomputer Fugaku, which RIKEN provided through the HPCI System Research Project (Project ID: hp220174). 
The simulation was also performed on Sakura, Cobra, and Raven clusters at the Max Planck Computing and Data Facility and the Cray XC50 at CfCA of the National Astronomical Observatory of Japan.

\section*{Data Availability}
The data that support the findings of this study are available upon request from the authors.


\bibliography{main} 
\bibliographystyle{aasjournal}


\appendix
\section{Data structure}
\label{appendix}
\subsection{Environment data}
\label{appendix sub: environment data}
The simulation domain of the initial conditions based on \citet{kiuchi_self-consistent_2023} comprises 13 levels of a fixed mesh refinement (FMR) embedded in a Cartesian geometry.
The finest FMR domain has the size of $L\in[-37.875,37.875]~\km$ with the grid spacing of $\Delta x= 0.15~\km$, while the coarsest has a grid size of $L\in[-1.55136,1.55136] \times 10^{5}~\km$ with a grid spacing of $\Delta x=614.4~\km$. 

To import the initial conditions data into {\textsc{pluto}} and run the simulation, we proceeded as follows:
\begin{enumerate}
\item for each refinement level, we create a cylindrical average around the $z-$axis for each cell, obtaining a 2D cylindrically symmetric Cartesian grid. To run the code in 2.5D we retain the $v_\phi$ components (which are also averaged);
\item we merge the 13 refinement levels into one, interpolating linearly between the regions where the grid resolution doubles in the original simulation grid;
\item we import the data into {\textsc{pluto}} creating a mock grid file that reflects the actual center of each cell;
\end{enumerate}

After the merger, we imported the initial conditions at $\td = 0.1~\s$ and $\td = 1~\s$. 
The first snapshot at $0.1~\s$ represents a case where the ejecta did not expand too further into the dense circumbinary environment ($\rho_\mathrm{floor} = 0.167~\g/\cm^3$) of the simulation box. 
The ejecta outer radius is about $~2\times 10^4~\km$ from the center of the simulation box; the second snapshot represents the system at the end of the original simulation, with the outer radius at $1.2\times 10^5~\km$ from the center. 

\subsection{Computational grid}
\label{appendix sub: computational grid}
For the simulation in {\textsc{pluto}} we use a grid of size $3549 \times 5541$ cells, with the radial cylindrical coordinate\footnote{Throughout the paper $r$ is used for the 2D cylindrical radius while $R$ stands for the 3D radius.} 
extending within the range $r = [0,1.5\times 10^5]\,\km$ and the vertical coordinate extending within the range $ z= [z_0, 1.5\times 10^5]\,\km$. \footnote{For the injection with $\theta_{\jet} = 10^\circ$ we use $z_0=50~\km$, since $r_\jet = z_0 \times \theta_\jet = \const$}. 
Our choice of a very low initial height to inject the jet is dictated by the fact that we want to explore the effect of the collimation in the innermost region of the ejecta, which retains most of the cavity. We start our jet to a distance comparable to the light cylinder radius of the compact object ($\simeq 40~\km$).

We use a combination of a uniform and a non-uniform mesh grid in $r-z$ coordinates with a decreasing resolution from the inner region of the simulation box to the outer boundaries. 
The grid mesh is uniform in the inner part to maintain a high resolution of the jet injection and the formation of the resulting high-pressure cocoon.
The uniform mesh has  $1000 \times 4000$ grid points extending in the ranges $r = [0, 100]~\km$ and $ z=[z_0, 5000]~\km$ with a resolution along both coordinates of $\Delta (r,z)_\mathrm{unif.} = [0.1, 1.2375]~\km $, allowing enough mesh coverage for the nozzle region.
Next to the uniform mesh, we placed a stretched mesh with $2549 \times 1541$ grid points extending along both coordinates within the range $(r,z) = [100,1.5\times 10^5] \times [5000, 1.5\times 10^5]~\km$ with a stretching ratio of $\simeq 1.0033$.
The number of grid points for this mesh is chosen such that its initial grid spacing is the same as the adjacent uniform mesh $\Delta (r,z)_\mathrm{s, init} = \Delta(r,z)_\mathrm{unif.} = 0.1\km$ and its final grid spacing is $\Delta(r,z)_\mathrm{s, final} = 500\,\km $. 
This ensures a smooth resolution increase without jumps for the entire simulation grid.

After switching off the jet injection, we lower the resolution of our computational domain using a $r \times z$ grid with a combination of uniform and geometrically stretched patches. The uniform grid $1000\times 900$ grid extends in the range $ (r,z) = [0, 5000]~\km \times [z_0, 5000]~\km $, with a resolution $\Delta x = 5~\km$, while the geometrically stretched grid with $1352^2$ cells extends in the range $(r, z) = [5000, 1.5\times 10^5]~\km$ in both directions. 

\section{Detailed jet setup and energy calculation}
\label{appendx: jet and energy}
\subsection{Jet injection}
\label{appendix sub: jet injection}
The jet is characterized by four parameters: luminosity $L_\jet$, opening angle $\theta_\jet$, enthalpy $h_\jet$, and the jet engine duration $\te$. 
In our setup, we inject the jet cylindrically through a nozzle such that the initial Lorentz factor $\Gamma_{0,\jet}$ is related to the initial choice of the jet opening angle via $\Gamma_{0,\jet} \simeq 1/(1.4 \theta_\jet)$ \citep{mizuta_opening_2013}. 
The following stress-energy tensor gives the energy density of a relativistic, unmagnetized fluid:
\begin{equation}
\label{eq: T00}
    T^{\mu\nu} = \rho h c^2 u^\mu u^\nu + P g^{\mu\nu}
\end{equation}
where $u^\mu$ is the 4-velocity and $g^{\mu\nu}$ is the metric tensor (which coincides with the Minkowski space-time in our case). In the laboratory frame, where $u^\mu = \Gamma(1,\vec{\beta})$, the energy flux of the fluid along z assumes the form \footnote{Some authors, like \citet{tominaga_aspherical_2009}, subtract the rest energy density of the fluid measured in the lab frame from this quantity.}:
\begin{equation}
    T^{0z} = \rho h c^2 \Gamma^2 \beta_z \quad . 
\end{equation}
The jet is injected through a nozzle with radius $r_\jet$ and area $\Sigma_{0z} \equiv \Sigma_\jet = \pi r^2_\jet$. 
At the nozzle, we inject a flux of energy per unit time such that:
\begin{equation}
\label{eq: jet_luminosity}
    L_\jet = \int_0^{r_\jet}  T_\jet^{0z} \de \Sigma_{0z} = \rho_\jet h_\jet c^2 \Gamma_{0,\jet}^2 \beta_{0,\jet} \Sigma_{\jet} \ ,
\end{equation}
where we imposed a top-hat profile of the jet (i.e., the properties of the jet do not vary with the position within the nozzle radius).
From Eq.~\eqref{eq: jet_luminosity}, we can derive the value for $\rho_\jet$. 
The expression of the enthalpy links pressure and density:
$h = 1 + \epsilon + P/\rho$, where $\epsilon(P,\rho)$ is the internal specific energy density of the fluid. The explicit expression for $\epsilon(P,\rho)$ depends on the choice of the EOS. In the case of a perfect fluid, we have $\rho \epsilon = P /(\gamma-1)$, where the adiabatic factor $\gamma = 4/3$ for a relativistic fluid, such that $\rho h c^2 = \rho c^2 + 4 P$. In the  case of a Taub-Matthews EOS for a Synge gas, we use the approximation showed in \citet{mignone_equation_2007} where the product $\rho c^2 h$ is expressed by

\begin{equation}
\label{eq: rhohc}
    \rho h c^2  = \dfrac{5}{2} P + \sqrt{\rho^2 c^4 + \dfrac{9}{4} P^2}
\end{equation}
Inverting Eq.~\eqref{eq: rhohc} and since we fix the enthalpy when we inject the jet, we find the following expression for the pressure:
\begin{equation}
    P = \dfrac{1}{8} \left(5 h - \sqrt{9 h^2 + 16} \right) \rho c^2 = h_\mathrm{eff} \rho c^2
\end{equation}
which reduces to $\rho h c^2 /4$ for a relativistic gas. 
Since we inject our jet vertically through the nozzle, the only non-zero component of $u^\mu$ is $\beta_z$. Since we fix the value of the luminosity $L_\jet$ and the enthalpy $h_\jet$ the quantities injected at the excision height $z_0$ and for $r \leq r_\jet$ are:
\begin{equation}
\label{eq: betar_jet}
    \beta_{r,\jet} = \beta_{\phi, \jet} = 0
\end{equation}
\begin{equation}
\label{eq: betaz_jet}
    \beta_{z, \jet} \equiv \beta_{0,\jet} =  \sqrt{1 - 1/\Gamma_{0,\jet}^2}
\end{equation}
\begin{equation}
\label{eq: rho_jet}
    \rho_\jet = \dfrac{L_j}{\Sigma_\jet h_\jet \Gamma^2_{0,\jet}\beta_{z, \jet} c^3}
\end{equation}
\begin{equation}
\label{eq: P_jet}
    P_\jet = h_{\mathrm{eff}} \rho_\jet c^2 
\end{equation}
Eqs.~(\ref{eq: betar_jet})--(\ref{eq: P_jet}) describe the jet injection at the nozzle and are constant in time by the entire duration of the engine activity as long as $L_\jet = \const$, otherwise $\rho_\jet$ and $P_\jet$ will vary according to the time dependency of $L_\jet$. 

Additionally, we enforce that, after $q$ time steps, the increase in the total energy over time follows the temporal law we specified for the jet luminosity. To do so, we measure the total energy added to the system up to that point on top of the ejecta energy of the initial conditions (i.e., $E_\mathrm{added} = E_\mathrm{measued} - E_\ej)$, and we compare it to the total energy we expect up to that moment integrating the luminosity function of our choice, i.e. $E(t) = \int_0^t L_\jet (t) \de t$. For a constant luminosity, the energy added to the system is trivial $E(t) = L_\jet \times t$.
If the energy in the system does not match the expected energy, we re-adjust iteratively the value of the jet pressure (and thus the jet density) at the injection point according to the following scheme:
\begin{equation}
    P_{\jet, n+q} = \begin{cases}
        (1+x) P_{\jet, n} & \textrm{if~$E_\mathrm{added} < E(t) $} \\
        P_{\jet, n} & \textrm{if~$E_\mathrm{added} = E(t) $} \\
        P_{\jet, n} / (1+x) & \textrm{if~$E_\mathrm{added} > E(t) $} \\
    \end{cases}
\end{equation}
where $P_{\jet, n}$ is the jet pressure at the nozzle at the current time step $n$, while $n+q$ indicates the $q$-th time step after $n$. The quantity $x = 10^{-3}$ is a small correction we give to the pressure at each time step to match the desired total energy of the system up to that point. In our simulations, we found that $q=500$ was sufficient to control the pressure value over time.

\subsection{Energy calculation}
Eq.~\eqref{eq: T00} expresses the fluid's total energy density. If we plug the definition of specific dimensionless enthalpy $h = 1 + \epsilon/ c^2 + P/\rho c^2$, we can rewrite it as:
\begin{equation}
\label{eq: total_energy}
    T^{00} = \rho c^2 \Gamma^2 + (\rho \epsilon+P) \Gamma^2 - P = D c^2 (\Gamma - 1) + D c^2 +  (\rho \epsilon+P) \Gamma^2 - P
\end{equation}
where we introduced the laboratory density $D = \rho \Gamma$. Writing the total energy density in this way allows us to distinguish  different contributions:
\begin{equation}
\label{eq: ekin}
    e_{\mathrm{kin}} = D c^2 (\Gamma -1)
\end{equation}
\begin{equation}
\label{eq: eint}
    e_{\mathrm{int}} =  (\rho \epsilon+P) \Gamma^2 - P = (h -1) \rho c^2 \Gamma^2 - P = \left(\dfrac{\gamma}{\gamma - 1} \Gamma^2 - 1\right) P 
\end{equation}
which are the kinetic and internal energy of the fluid in the lab frame, respectively. In case of a Taub-Matthews EOS we can still find an effective adiabatic factor $\gamma$ that allows writing Eq.~\eqref{eq: eint} in the form shown and, generalizing, we can write $\gamma/(\gamma -1) = (h -1) \rho c^2 / P$ \citep{mignone_equation_2007}. 
Integrating Eqs.~\eqref{eq: ekin} and \eqref{eq: eint} over the entire volume gives the total energy of the relativistic fluid in the simulation measured in the laboratory frame. 

\section{Jets with same terminal velocity}

\begin{figure}
    \centering
    \includegraphics[width=1\linewidth]{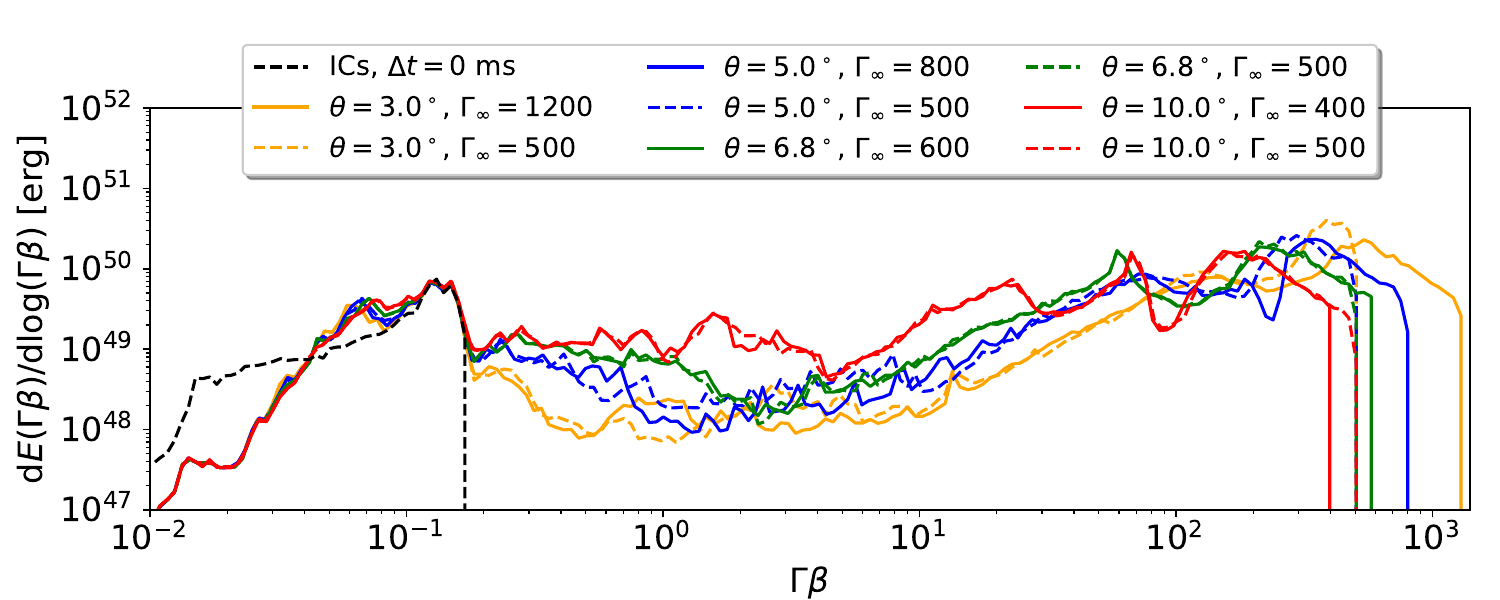}
    \vskip -0.2cm
    \caption{Differential energy distribution with respect to the proper velocity $\Gamma \beta $ of relativistic jets with $L_\jet = 10^{51}~\erg/\s$ and $\theta_\jet = [3,5, 6.8, 10]^\circ$ after $300~\millis$ of injection. The solid lines represent runs with $h_\jet = 100$ and $\Gamma_{\jet,0} = [12,8,6,4]$, respectively, such that the final Lorentz factor $\Gamma_\infty = \Gamma_{\jet,0} h_\jet$ is not fixed. The dashed lines represent runs with the same opening angles but fixed $\Gamma_\infty = 500$. }
    \label{fig: energy_vs_velocity_gamma_infty}
\end{figure}

\begin{figure}
    \centering
    \includegraphics[width=1\linewidth]{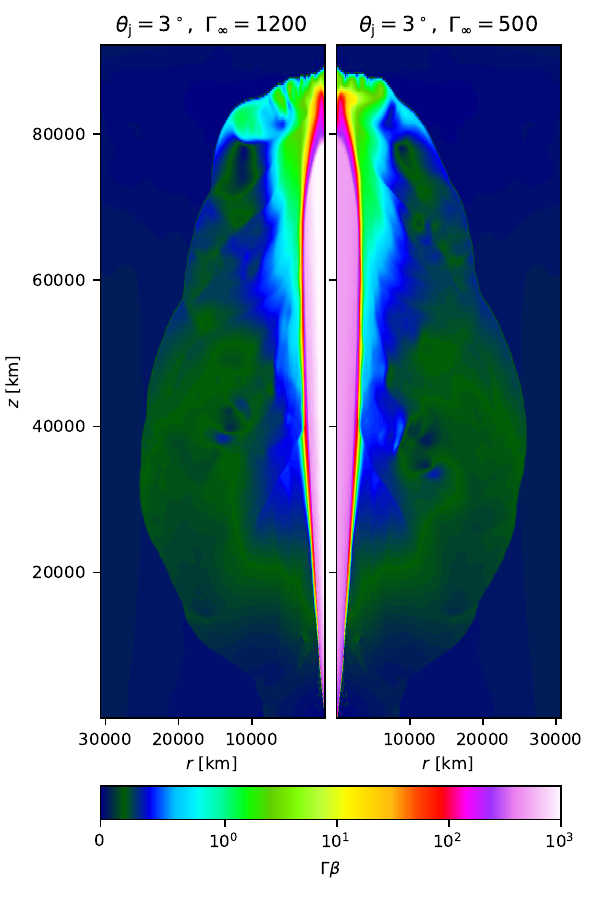}
    \vskip -0.3cm
    \caption{Comparison, after $t_\jet = 300~\millis$, between the velocity maps of two jets with the same opening angle $\theta_\jet = 3^\circ$, $L_\jet = 10^{51}~\ergsec$, but different $\Gamma_\infty$. We modified the color map and scale with respect to the other velocity maps to enhance the difference between the two runs. {Notice the similarity in the overall structure.}} 
    \label{fig: jet_different_gamma_infinity}
\end{figure}

The previous runs we presented for different opening angles were initialized so that the initial opening angle $\theta_\jet$ determines the initial Lorentz factor via $\Gamma_{0,\jet} \simeq 1/(1.4\theta_\jet) \simeq [8, 4]$ (for $\theta_{0,\jet} = [5^\circ, 10^\circ]$, respectively), and keeping the enthalpy of the jet fixed at $h_\jet = 100$. Since $\Gamma_\infty \simeq \Gamma_{0,\jet} h_\jet$, the terminal Lorentz factor of the system depends on the initial Lorentz factor and the specific enthalpy used to initialize the jet.
We show here runs where we keep the terminal Lorentz factor constant, varying the specific enthalpy according to the choice of the initial Lorentz factor, i.e., using $ h_\jet = \Gamma_\infty / \Gamma_{0,\jet}$, where $\Gamma_\infty = 500$, which is value high enough to avoid cold jets. 
We run three cases with the same luminosity, fixed at $L_\jet = 10^{51}~\erg/\s$, and different initial opening angles $\theta_{\jet} = [3, 5, 6.8, 10]^\circ$. 
The choice of the angles is such that they correspond to initial Lorentz factors of $\Gamma_{\jet,0} = [12,8,6,4]$, respectively.
In Fig.~\ref{fig: energy_vs_velocity_gamma_infty}, we show the differential distribution of the energy with respect to the proper velocity where we compared the cases with {$h_\jet = 100$ (solid lines) with the cases with $h_\jet = \Gamma_\infty / \Gamma_{0,\jet}$ where $\Gamma_\infty = 500$ (dashed lines)}. 
We notice that the energy distribution is essentially the same between the two types of runs except for the ending portion of the tail, where energy accumulates. 

In Fig \ref{fig: jet_different_gamma_infinity}, we show two jets with the same luminosity, opening angle $\theta_\jet = 3^\circ$ but different terminal velocity (i.e., different enthalpy). We chose a small opening angle to emphasize the difference in $u_\infty = 1200$ vs.~500, which translates into $h_\jet = 100$ vs.~$\simeq 42$. Both jets are hot enough, so most of the initial energy is injected into the thermal channel. The jet expansion, head velocity, and drilling through the ejecta are very similar in both cases. We conclude that choosing a fixed $\Gamma_\infty$, as long as the jet is hot, does not affect our results and runs.

\section{Afterglow modeling}
\label{appendix: afterglow model}
\subsection{Framework}
The synchrotron emission of the GRB afterglow is modeled after \citet{sari_spectra_1998-1} and \citet{granot_images_1999}. 
 
The received flux is given by the volume integral:
\begin{equation}
    \begin{aligned}
        \mathcal{F}_\nu (T, \theta_\obs) &= \left. \dfrac{1}{4\pi D^2}\int_0^{2\pi}\de \phi \int_0^{\theta_\max} \de \theta \sin\theta \dfrac{R^2 \Delta R ~ \epsilon'_{\nu'}}{\Gamma^2 (1-\beta\mu)^2} \right|_{t=T+\mu R/c} \\
        & \simeq \left. \dfrac{1}{4\pi D^2}\int_0^{2\pi} \int_0^{\theta_\max} \dfrac{\de \phi \de \theta \sin\theta R^3 ~ \epsilon'_{\nu'}}{12 \Gamma^4(1-\beta_\mathrm{sh}\mu)(1-\beta\mu)^2} \right|_{t=T+\mu R/c} 
    \end{aligned}
\end{equation}
where we assumed that the observed radiation is produced by a thin shell of width $\Delta R \simeq R / [12 \Gamma^2 (1-\beta_\mathrm{sh} \mu)]\ll R_\mathrm{sh}$, smaller than the size of the shock radius. In the formula, $D$ is the distance from the source, $\epsilon'_{\nu'}$ is
the comoving frame emissivity, and $\mu(\theta,\phi)$ is the angle between observer direction and local velocity, defined as:
\begin{equation}
    \mu = \cos\theta \cos \theta_\obs + \sin\theta \sin\theta_\obs \cos\phi
\end{equation}
Furthermore, the integral is calculated over different emission times, which depend on the position $(\theta,\phi)$. 
The implicit formula
\begin{equation}
\label{eq: t_lab}
    t = T + \dfrac{\mu}{c} R(t)
\end{equation}
relates the laboratory time $t$ at each position $(\theta,\phi)$ to a given observer time $T$. Solving it we obtain $t = t(T,\theta,\phi)$ for  fixed $\theta_\obs$. The functional form of $R(t)$ is found taking some approximations: we assume that the jet expands in a cold and homogeneous interstellar medium; 
the jet is assumed to be axisymmetric, the energy of the blast wave is angle-dependent, and each segment expands spherically with a portion of the total energy given by the local isotropic equivalent energy $E_\Omega \equiv \de E (\theta)/\de \Omega = E_0 f(\theta)$.

\begin{figure}
    \centering
    \includegraphics[width=1\linewidth]{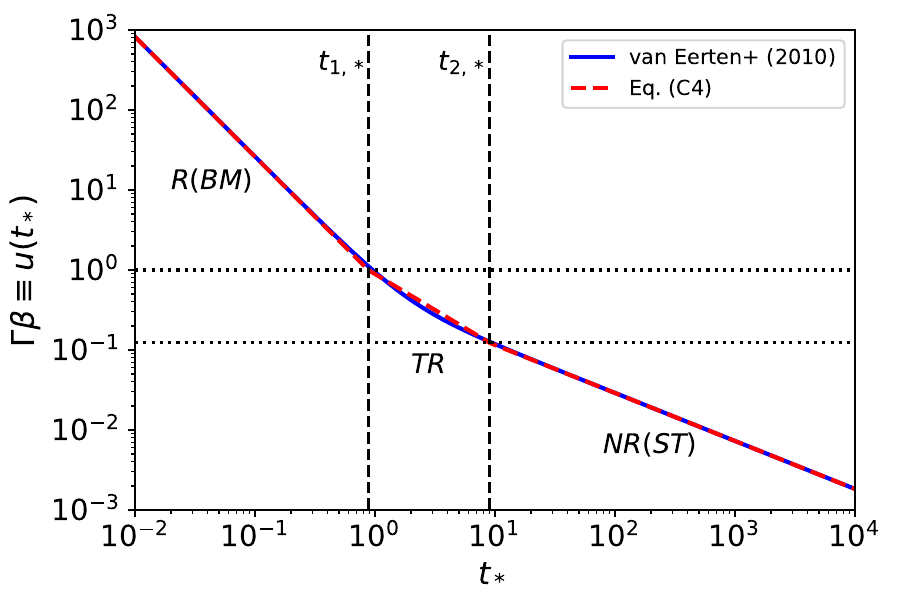}
    \vskip -0.2cm
    \caption{Comparison between the formula used in \citet{van_eerten_off-axis_2010} for the shock proper velocity and our broken power law expressed by Eq.~\eqref{eq: gamma_beta}. The curves are plotted in dimensionless time $t_*$ and are scale-free. The dashed vertical lines indicate the two transition times $t_1$ and $t_2$ between the relativistic, trans-relativistic, and the trans-relativistic and Newtonian regimes, respectively. The horizontal dotted lines represent the values of $u_\sh$ at the transition times. Our formula interpolates the two asymptotic regimes for $t_* \rightarrow 0$ and $t_* \rightarrow \infty$ with a softer power law between the two with $u \propto t^{-0.9}$.}
    \label{fig: proper_velocity_comparison}
\end{figure}

Each shock segment is described by an ultra-relativistic blast wave, which gradually decelerates. We distinguish three regimes: i) R: \textit{relativistic} ($\Gamma \beta > 1$), in which the evolution of the (proper) velocity of the shock is described by the self-similar solution of \citet{blandford_fluid_1976}; ii) TR: \textit{trans-relativistic} ($0.1 \lesssim \Gamma\beta \leq 1 $), which marks the gradual passage from a relativistic blast wave to a Newtonian one and. iii) NR: \textit{non-relativistic} ($\Gamma\beta \lesssim 0.1 $), in which the shock dynamics is described by the Sedov-Taylor solution \citep{taylor_formation_1950, Sedov_1959}. 

Unlike \citet{van_eerten_off-axis_2010}, we do not combine the two self-similar solutions by summing their squares under the square root. Instead, we use a broken power law for the proper velocity to describe the three phases of the shock radius evolution.
The self-similar solution for $\Gamma\beta$ reads
\begin{equation}
\label{eq: gamma_beta}
    (\Gamma\beta)_\sh \equiv u_\sh = \begin{cases}
        C_\BM t^{-3/2} & \quad \mathrm{(R)}, \quad  \ \ \mathrm{if} \ \ \Gamma\beta > 1 \\
        C_\BM^{3/5} t^{-9/10} & \quad \mathrm{(TR)}, \quad \mathrm{if} \ \ 0.12 \lesssim \Gamma\beta \leq 1 \\
        C_\ST t^{-3/5} & \quad \mathrm{(NR)}, \quad \mathrm{if} \ \ \Gamma\beta \lesssim 0.12 \\
    \end{cases}    
\end{equation}
where 
\begin{equation}
\label{eq: coefficients}
    C_\BM = \sqrt{\dfrac{17 E_\Omega}{8 \pi n_0 m_\mathrm{p} c^5}} \quad , \quad C_\ST = \dfrac{2}{5} \times 1.15 \left( \dfrac{E_\Omega}{n_0 m_\mathrm{p} c^5}\right)^{1/5}
\end{equation} 
are the coefficients of the Blandford-McKee and the Sedov-Taylor self-similar solutions, respectively.

\begin{figure*}
    \centering
    \includegraphics[width=1\linewidth]{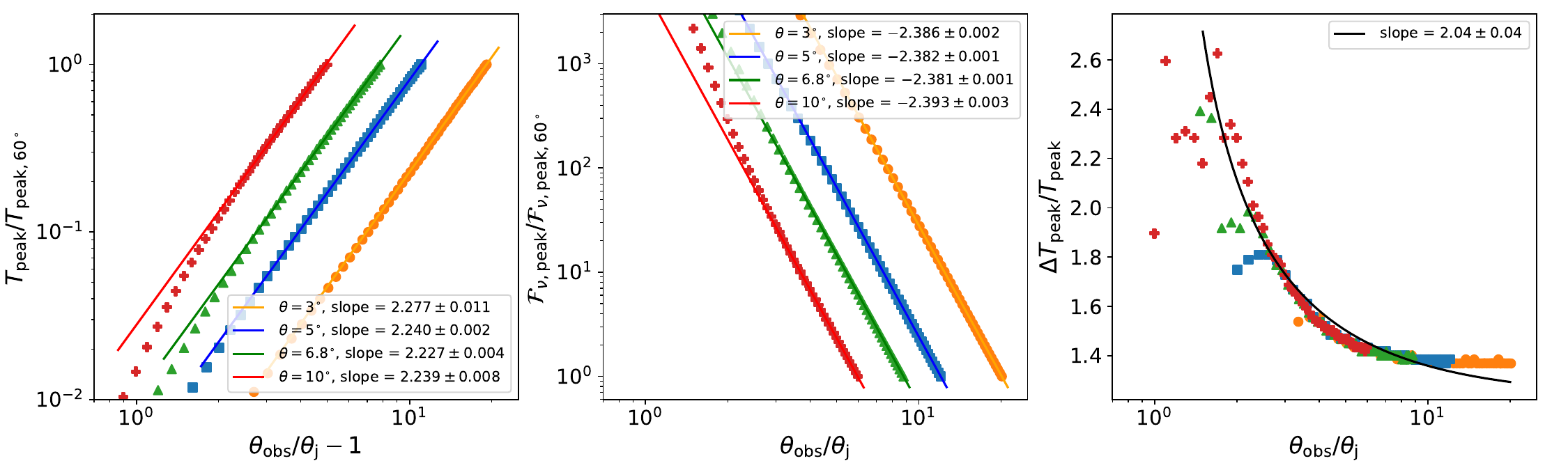}
    \vskip -0.2cm
    \caption{Peak position, peak height and peak spread of the afterglow curves as a function of the viewing angle $\theta_\obs$, for four different opening angles: $3^\circ$ (blue points), $5^\circ$ (orange), $6.8^\circ$ (green), $10^\circ$ (red). We superposed to the points the power-law fitting lines with the same color as the points. The viewing angle is normalized with respect to the jet opening angle of the corresponding run. In contrast, $T_\peak$ and $\mathcal{F}_{\nu,\peak}$, for simplicity, are normalized with respect to their values for $\theta_\obs=60^\circ$ (the last point of the series).}
    \label{fig: theta_obs_scaling}
\end{figure*}

The trans-relativistic (TR) regime is obtained by modifying the Blandford-McKee solution raised to the power of $3/5$, which imposes a different slope to make the transition to a non-relativistic regime smoother.
The transition times between the different regimes are found imposing $(\Gamma\beta)_\R = (\Gamma\beta)_\TR$, and $(\Gamma\beta)_\TR = (\Gamma\beta)_\NR$. We find
\begin{equation}
    t_{\R \rightarrow \TR} = C_\BM^{2/3} \quad , \quad t_{\TR\rightarrow \NR} = C_\BM^{2} C_\ST^{-10/3}
\end{equation}
We can rewrite the formulas more concisely, introducing a characteristic time scale for the blast wave segment such that
\begin{equation}
    t = \alpha t_*  \quad , \quad  R = \alpha c R_* \ , 
\end{equation}
where
\begin{equation}
    \alpha(\theta) = \left( \dfrac{E_\Omega }{n_0 m_\mathrm{p} c^5} \right)^{1/3} = 337~\days \left(\dfrac{E_0 f(\theta)}{10^{51}~\erg}\right)^{1/3} \left(\dfrac{n_0}{1~\cm^{-3}}\right)^{-1/3} \ ,
\end{equation}
which depends on $\theta$ via the local isotropic energy. 
We can rewrite the coefficients in Eq.~\eqref{eq: coefficients} as $C_{\BM,*} = \sqrt{17/8\pi}$ and $C_{\ST,*} = 0.46$, and the transition times as:
\begin{equation}
    \begin{aligned}
        & t_{1,*} = t_{\R \rightarrow \TR, *} = \left( \dfrac{17}{8\pi}\right)^{1/3} \simeq 0.878 \quad , \\ 
        & t_{2,*} = t_{\TR\rightarrow \NR, *} = \dfrac{17}{8\pi} \left( \dfrac{2}{5} \times 1.15 \right)^{-10/3} \simeq 9 \ .
    \end{aligned}
\end{equation}
Substituting, these values in Eq.~\eqref{eq: gamma_beta} correspond to the values of the proper velocity:
\begin{equation}
    u_{\R \rightarrow \TR} = 1 \quad , \quad u_{\TR\rightarrow \NR} = C_{\ST}^3 C_\BM^{-6/5} \simeq 0.123
\end{equation}
A comparison between the proper velocity formula used by \citet{van_eerten_off-axis_2010} and our approximation is shown in Fig.~\ref{fig: proper_velocity_comparison}.

The Lorentz factor of the fluid is related to the Lorentz factor of the shock via the jump conditions, giving $u = u_\sh/\sqrt{2}$ in the relativistic and trans-relativistic regimes, and $u = 3 u_\sh/4$ in the non-relativistic regime.
The shock radius is given by
\begin{equation}
\label{eq: int_R}
    R_\sh(t) = c \int_0^t \beta_\sh(\tilde{t}) \de \tilde{t} \quad \longrightarrow \quad R_{\sh,*}(t_*) = \int_0^{t_*} \beta_\sh(\tilde{t})\de \tilde{t}
\end{equation}
with $\beta_\sh = u_\sh /\sqrt{1+u_\sh^2}$. For each regime, the integration is analytical and yields to 
\begin{equation}
\label{eq: radius_time}
\begin{aligned}
    & R_{\sh, \R,*} = {}_2F_1 \left( \dfrac{1}{3} , \dfrac{1}{2}, \dfrac{4}{3} ; - \dfrac{t_*^3}{C_\BM^2}\right) t_* , \\ 
    & R_{\sh, \TR,*} = 10 C_\BM^{3/5} ~ {}_2F_1\left(-\dfrac{1}{18}, \dfrac{1}{2}, \dfrac{17}{18}, -\dfrac{C_\BM^{6/5}}{t_*^{9/5}}\right) t_*^{1/10} \\
    & R_{\sh, \NR,*}  = C_\ST t_*^{2/5}
\end{aligned}
\end{equation}
where ${}_2F_1(\dots)$ is Gauss' hypergeometric function. The integral in Eq.~\eqref{eq: int_R}  can be broken into the individual contributions of the terms expressed by Eq.~\eqref{eq: radius_time} for the different regimes:
\begin{equation*}
    R_{\sh,*} = 
    \begin{cases}
        R_{\sh, \R,*} (t_*)  & t_* \leq t_{1,*} \\ 
        R_{\sh, \R,*} (t_{1,*}) + \left. R_{\sh, \TR,*} \right|^{t_*}_{t_{1,*}} & t_{1,*} < t_* \leq t_{2,*} \\
        R_{\sh, \R,*} (t_{1,*}) + \left. R_{\sh, \TR,*} \right|^{t_{2,*}}_{t_{1,*}} + \left. R_{\sh, \NR,*} \right|^{t_*}_{t_{2,*}}& t_{2,*} < t_*
    \end{cases}
\end{equation*}
where $\left. x(t) \right|^{b}_{a} = x(b)- x(a)$. Substituting the constants, we find:
\begin{equation}
    R_{\sh,*} = 
    \begin{cases}
        R_{\sh, \R,*} (t_*)  & \ \ \ \qquad t_* \leq t_{1,*} \\ 
        R_{\sh, \TR,*} (t_*)  - 8.176 & t_{1,*} < t_* \leq t_{2,*} \\
        R_{\sh, \NR,*} (t_*) + 1.8 & t_{2,*} < t_*
    \end{cases}   
\end{equation}
This method helps to quickly find the numerical root for $t$ in Eq.~\eqref{eq: t_lab}, which can be easily rewritten as $t_* = T_* + \mu R_*(t_*)$, with $T_* = T/\alpha$. When the scale $\alpha(\theta)$ becomes very small (i.e. for $\theta \gg \theta_\jet$ and $f(\theta) \ll 1$), then, at later times, $T_*$ becomes larger than $R_*(t_*)$ and we can impose $t \simeq T$ to speed up the calculations.

The shocked medium is characterized by its comoving number density $n'$, internal energy density $e_i'$, and magnetic field strength $B'$. Those are given as follows 
 \citet{sari_spectra_1998-1}:
\begin{equation}
    n' = 4 \Gamma n_0
\end{equation}
\begin{equation}
    e_i' = (\Gamma -1) n' m_\mathrm{p} c^2 = 4 \Gamma (\Gamma -1) n_0 m_\mathrm{p} c^2
\end{equation}
\begin{equation}
    B' = \sqrt{8\pi \epsilon_\mathrm{B} e_i'}
\end{equation}
where $\epsilon_\mathrm{B}$ is a parameter that expresses the energy fraction of the shocked matter going into the magnetic field.

The non-thermal electrons generating the synchrotron emission are described by an energy spectrum with a single power-law $p$, and a constant fraction $\epsilon_\mathrm{e}$ of the shock energy is assumed to go into these accelerated electrons. 
The synchrotron spectrum is approximated by a broken power law, with a characteristic frequency $\nu'_\mathrm{m}$ and a cooling frequency $\nu'_\mathrm{c}$. 
These are given by:
\begin{equation}
    \nu_\mathrm{m}' = \dfrac{3}{16}\left[ \epsilone \left(\dfrac{p-2}{p-1}\right) \dfrac{\mprot}{\me} (\Gamma-1)\right]^2 \dfrac{\qelectron B'}{\me c}
\end{equation}
\begin{equation}
    \nu_\mathrm{c}' = \dfrac{3}{16}\left[ \dfrac{3 \me c \Gamma}{4 \sigma_\mathrm{T} \epsilonB e_i' t}\right]^2 \dfrac{\qelectron B'}{\me c}
\end{equation}
where $\qelectron$ is the electron charge and $\sigma_\mathrm{T} = 8\pi \qelectron^4 / (3 \me^2 c^4)$ is the Thomson cross section.
The emissivity $\epsilon'_{\nu'}$, in case of slow cooling ($\nu'_\mathrm{m} < \nu'_\mathrm{c}$) is given by:
\begin{equation}
    \epsilon'_{\nu'} = \epsilon'_{\nu', \mathrm{peak}} \begin{cases}
        (\nu'/ \nu'_\mathrm{m})^{1/3} \quad & \nu' < \nu'_\mathrm{m} \\
        (\nu'/\nu'_\mathrm{m})^{(1-p)/2} \quad & \nu'_\mathrm{m} \leq \nu' < \nu'_\mathrm{c} \\
         (\nu'_\mathrm{c}/\nu'_\mathrm{m})^{(1-p)/2}  (\nu'/\nu'_\mathrm{c})^{-p/2} \quad & \nu'_\mathrm{c} \leq \nu'
    \end{cases}
\end{equation}
where $\epsilon'_{\nu', \mathrm{peak}}$ is the peak emissivity \citep{granot_images_1999}:
\begin{equation}
    \epsilon'_{\nu', \mathrm{peak}} \simeq 0.88 \times \dfrac{256}{27} \left( \dfrac{p-1}{3p -1} \right) \dfrac{q_\mathrm{e}^2}{m_e c^2} n' B' \ .
\end{equation}
Finally, the rest-frame frequency $\nu'$ is connected to the observer frequency $\nu$ by:
\begin{equation}
    \nu' = \Gamma ( 1- \beta\mu) \nu 
\end{equation}

\begin{figure*}
    \centering
    \includegraphics[width=1\linewidth]{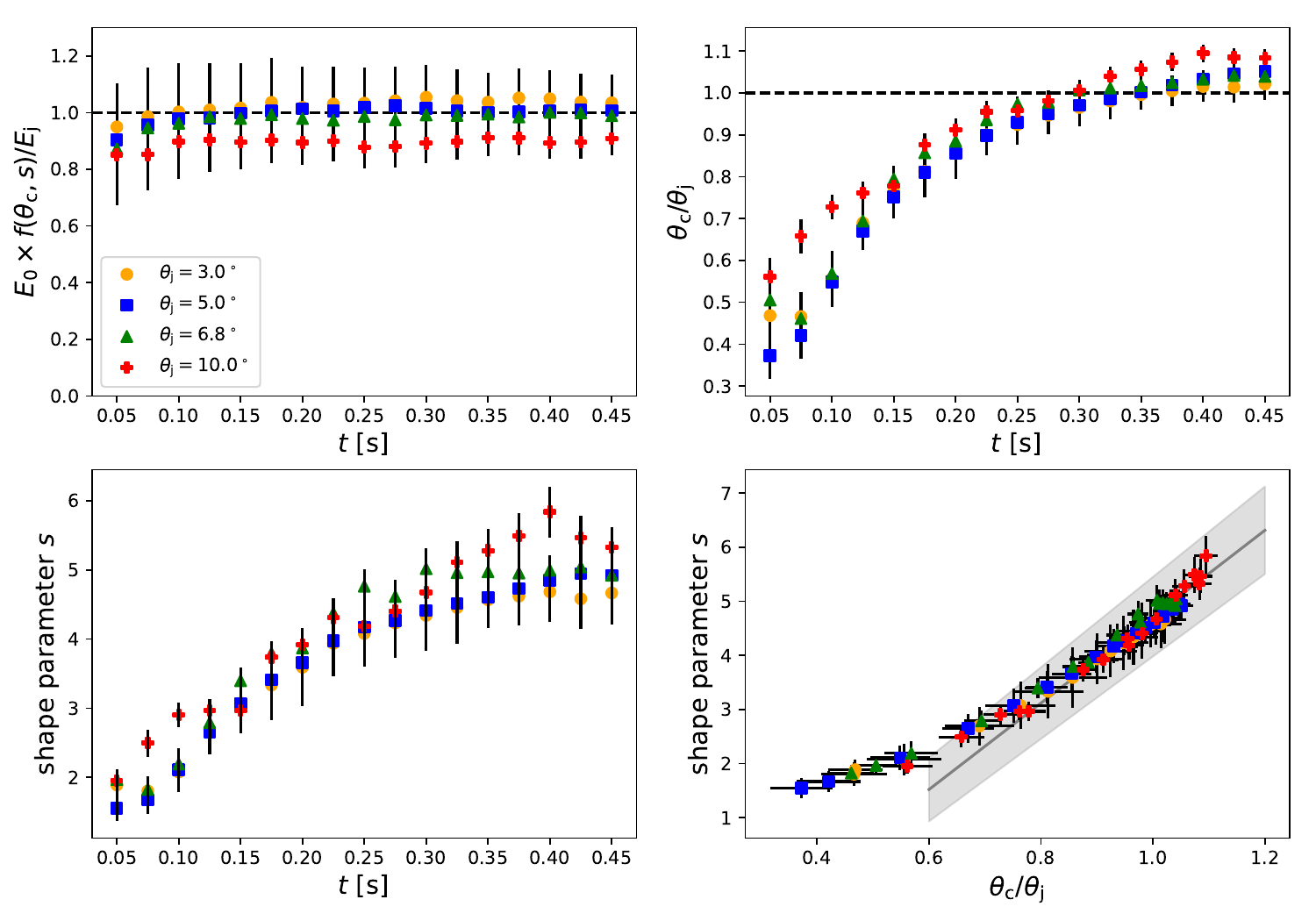}
    \vskip -0.3cm
    \caption{Time evolution of the parameters of the generalized Gaussian fit for four jet simulations with $\theta_\jet = [3,5,6.8,10]^\circ$ and $L_\jet = 10^{51}~\ergsec$. A specific color and marker identify each run in each plot. The top left panel shows $E_0$ multiplied by an integration factor that depends upon $\theta_\mathrm{c}$ and the shape parameter normalized to the jet energy $E_\jet$, namely $f(\theta_\mathrm{c}, s) = 2 \pi \theta_\mathrm{c}^2 [ c_1(s) - c_2(s) \theta_\mathrm{c}^2]$. The top right panel shows the ratio $\theta_\mathrm{c}/ \theta_\jet$ vs time. Analogously, the bottom left panel shows the shape parameter $s$ vs.~time, and the bottom right parameter shows the linear correlation between $\theta_\mathrm{c}$ and the shape parameter.}
    \label{fig: gaussian_data_evolution}
\end{figure*}

\subsection{Degeneracy relations}
The afterglow light curves are subject to degeneracy relations for the peak position and the peak height, derived from the following scaling laws \citep{nakar_detectability_2002, nakar_afterglow_2020}:
\begin{equation}
\label{eq: peak_scaling}
    T_\peak \propto E_0^{1/3} n_0^{-1/3} (\theta_\obs - \theta_\jet)^{2}
\end{equation}
\begin{equation}
\label{eq: flux_scaling}
    \mathcal{F}_{\nu,\peak} \propto E_0 n^{\frac{p+1}{4}} \epsilone^{p-1} \epsilonB^{\frac{p+1}{4}} \theta_\obs^{-2p} \nu^{\frac{1-p}{2}} D^{-2}
\end{equation}
for a fixed observer time $T$. Here, $E_0 = E_\Omega(\theta =0)$. Additionally, the scaling for $\theta_\obs$ becomes more precise as long as $\theta_\obs \gg \theta_\jet$.
We verified the scaling relations by numerically integrating the light curves for several viewing angles, from $\theta_\obs = 14^\circ$ to $60^\circ$ at steps of $1^\circ$, and for four different opening angles of the jet, i.e. $\theta_\jet = [3, 5, 6.8, 10]^\circ$. 
We show the results of the peak time and the peak flux as a function of the viewing angle in Fig.~\ref{fig: theta_obs_scaling}. From the fitting lines we see that, for a ratio $\chi = \theta_\obs/\theta_\jet$ sufficiently larger than unity, the scaling relation is a power law with slope $\simeq 2.2$ (instead of $2$ in Eq.~\eqref{eq: peak_scaling}) for $T_\peak$ and a slope of $\simeq -2.4 p$ for $\mathcal{F}_{\nu,\peak}$ (instead of $-2 p$ from Eq.~\eqref{eq: flux_scaling}). 
Furthermore, the afterglow peak tends to be wide for $\chi \simeq 1 - 3 $ and narrower for $\chi \gg 1 $ until the shape converges to a certain curve. 
Combining Eqs.~\eqref{eq: peak_scaling} and \eqref{eq: flux_scaling}, and taking into account the corrections above for the scaling with respect to $\theta_\obs$, we get the generalized scaling relations for the peak time and the peak flux:
\begin{equation}
\begin{aligned}
    \dfrac{p+5}{4} \log\left(\dfrac{\tilde{n}_0}{n_0}\right) + \dfrac{p+1}{4} \log\left(\dfrac{\tilde{\epsilon}_\mathrm{B}}{{\epsilonB}}\right) + (p-1) \log\left(\dfrac{\tilde{\epsilon}_\mathrm{e}}{{\epsilone}}\right) =  \\
    = (3 l_1 +  l_2 p) \log\left(\dfrac{\tilde{\theta}_\obs}{{\theta_\obs}}\right) + 3 l_1 \log\left(\dfrac{1-1/\tilde{\chi}}{{1-1/\chi}}\right)
\end{aligned}
\end{equation}
with $l_1 \simeq 2.2$ and $l_2 \simeq 2.4$. 
If the observer's angle does not change, the second term goes to 0, and the previous equation reduces to the scaling relation of \citet{takahashi_diverse_2021}.
We remark that, even if the peak time and peak flux coincide for different viewing angles, the previous equation does not define a family of perfectly superposing curves since the afterglow light curve tends to spread the more the observer gets closer to the jet's opening angle. 
This is visible from the third panel of Fig.~\ref{fig: theta_obs_scaling} where we measured the time $\Delta T$ between the peak time $T_\peak$ and the asymptotic decay of the light curve (i.e., the observer time when the flux decreases by roughly 10\% from its peak value), which roughly follows the power law scaling of a spreading jet \citep{granot_structure_2007}.

\subsection{Jets described by a generalized Gaussian and their time evolution}
\label{appendix: jet_shape}
The integral of the energy per solid angle of a jet described by Eq.~\eqref{eq: dEdOmega} is rather complicated and is analytical only for specific integer or fractional values of $s$; thus, it is better to approximate the integrated function using a Taylor series expansion of the result around $\theta_\mathrm{c}=0$, as it follows:
\begin{equation}
    \label{eq: int_total_energy}
    \begin{aligned}
    E_\jet &= 2 \pi E_0 \int_0^{\theta_\max} \de \theta \sin \theta \exp\left[- \dfrac{1}{2} \left(\dfrac{\theta}{\theta_\mathrm{c}} \right)^{s} \right] \\ &  = 2 \pi E_0  \theta_\mathrm{c}^2 ( c_1 - c_2 \theta_\mathrm{c}^2) + \mathcal{O}(\theta_\mathrm{c}^6)
    \end{aligned}
\end{equation}
Expressing $\sin\theta$ as a Taylor series around $\theta_\mathrm{c} =0$ and integrating term-by-term (assuming that $\theta_\max \gg \theta_\mathrm{c}$) we find the following expression for the integral:

\begin{equation}
    E_\jet (\theta_\mathrm{c},s) =  \dfrac{2\pi E_0 \theta^2_\mathrm{c}}{s} \sum_{n=0}^\infty (-1)^n \dfrac{\theta_\mathrm{c}^{2n} 4^{\frac{n+1}{s}}}{(2n+1)!} \Gamma\left( \dfrac{2n+2}{s}\right) , 
\end{equation}
where $\Gamma(x)$ is Gauss' Gamma function in this context.
We find that the first two terms $c_1(s)$ and $c_2(s)$ have the following exact forms: 
\begin{equation}
    \label{eq: coefficients_gaussian}
    c_1(s) = \dfrac{4^{1/s}}{s} \Gamma\left(\dfrac{2}{s}\right) \quad , \quad c_2(s) = \dfrac{4^{2/s}}{6 s} \Gamma\left(\dfrac{4}{s}\right)  \ .
\end{equation}
For $s \rightarrow \infty$, the edges of the jet become infinitely sharp, and the Gamma functions of the coefficients become $\simeq s/(2n+2)$. The jet shape degenerates into a Heaviside step function $E_0 \Theta(\theta_c - \theta)$, thus the integral of Eq.~\eqref{eq: int_total_energy}, dropping a factor $2\pi$, will give exactly $1 - \cos(\theta_\mathrm{c}) = \sum_{j=1}^\infty (-1)^{j+1} \frac{1}{2j!}\theta_\mathrm{c}^{2j} = \frac{1}{2!} \theta_\mathrm{c}^ 2 - \frac{1}{4!} \theta_\mathrm{c}^4 + \ldots$, which are the asymptotic limits of the coefficients $c_1$ and $c_2$. 

In Fig.~\ref{fig: gaussian_data_evolution}, we show the time evolution of the fitting parameters for four values of the opening angle $\theta_\jet = [3,5,6.8,10]^\circ$ and fixed $L_\jet = 10^{51}~\ergsec$. 
We took snapshots from $25~\millis$ to $450~\millis$ of continuous jet injection, at $\Delta t= 25~\millis$ from each other. 
We cut off the rest of the slow ejecta in each snapshot and fitted the on-axis peak with a generalized Gaussian. 
We used Eqs.~(\ref{eq: int_total_energy}) and (\ref{eq: coefficients_gaussian}) to calculate the total energy under the fitting function and compared it to the jet energy for each given time. 
We see that for all the runs but one ($\theta_\jet = 10^\circ$), the result matches the expected jet energy very well. 
The discrepancy at larger angles can be explained considering that there is a non-negligible amount of energy outside the peak, thus giving less than expected energy. 
The value of $\theta_\mathrm{c}$ grows in the first 250 ms of injection to roughly stabilize around the value of the jet's opening angle $\theta_\jet$. Also, the shape parameter of the Gaussian seems to converge to a certain value at later times, around $5$. However, due to its wide angular spread, the run with $\theta_\jet = 10^\circ$ seems to jump to higher values (flatter central distribution). Not surprisingly, $\theta_\mathrm{c}$ and the shape parameters are linearly correlated at later times, as shown in the bottom-right panel of the figure. 
From these data, we can conclude that taking the energy distribution $E_\Omega$ too early might strongly affect the final result when the synthetic light curves are calculated since its parameters tend to stabilize at later times. 

\end{CJK}
\end{document}